\documentclass[12pt]{article}

\usepackage{amsmath,amssymb,latexsym,theorem,epsfig,amscd}
\usepackage[all]{xy}
\CompileMatrices

\newcommand{\be}{\begin{equation}}
\newcommand{\ee}{\end{equation}}

\newcommand{\rank}{{\text{ rank} }}
\newcommand{\effective}{{\text{effective} }}
\newcommand{\ch}{{\text{ch}\; }}

\newcommand{\td}{{\text{td}\; }}
\newcommand{\pt}{{\text{pt} }}

\newcommand{\cp}[1]{{\mathbb P}^{#1}}
\newcommand{\op}[1]{\operatorname{#1}}
\newcommand{\ocp}[1][]{{\mathcal O}_{\cp{1}}{#1}}
\newcommand{\oy}[1][]{{\mathcal O}_B{#1}}
\newcommand{\oz}[1][]{{\mathcal O}_{B^{'}}{#1}}
\newcommand{\ox}[1][]{{\mathcal O}_X{#1}}

\newcommand{\num}[1]{{\#}\text{\bfseries #1}}

\newcommand{\ses}[3]{\xymatrix{
0 \ar[r] & {#1} \ar[r] & {#2} \ar[r] & {#3} \ar[r] & 0 \\ }}
\newcommand{\les}[8]{\xymatrix{       &      & ...  \ar[r]  &  {#1}    \ar@{->} `r[d] `[l] `^dl[dlll]  `^dr/14pt[dll]    [dll] \\
&  {#2} \ar[r] & {#3} \ar[r] & {#4}  \ar `r/10pt[d] `[l]  `^dl[dlll]  `^dr/14pt[dll]   [dll] \\ 
& {#5} \ar[r]  & {#6} \ar[r] & {#7}  \ar `r/10pt[d] `[l]  `^dl[dlll]  `^dr/14pt[dll]   [dll] \\
&  {#8} \ar[r] & ... & & }}
\newcommand{\rs}[5]{
\xymatrix{
         &            &            &  0         & \\ 
         &            &            & {#5} \ar[u] & \\
0 \ar[r] & {#1} \ar[r] & {#2} \ar[r] & {#3} \ar[r] \ar[u] & 0 \\  
         &            &            & {#4} \ar[u]\ar@{.>}[ul] & \\
         &            &            &  0  \ar[u] &
}}

\newcommand{\ts}[4]
{\xymatrix{
0 \ar[r] & {#1} \ar[r] & {#2} \ar[r] & {#3} \ar[r] & 0 \\  
         &            & {#4} \ar[u] \ar@{.>}[ul]^{\gamma} \ar@{.>}[ur]_{\lambda}  &            & \\
         &            &  0  \ar[u] &            &
}}

\newtheorem{theo}{Theorem}[section]

{\theorembodyfont{\rmfamily} }
{\theorembodyfont{\rmfamily} \newtheorem{ex}[theo]{Example}}
\numberwithin{equation}{section}

\newcommand{\Appendix}[1]{%
  \refstepcounter{section}%
  \addcontentsline{toc}{section}%
    {\bfseries\appendixname~\thesection:\  #1}%
    {\medskip\noindent \Large\bfseries\appendixname\ \thesection:\ #1}%
\sectionmark{#1}\smallskip\noindent
\renewcommand{\theequation}{\thesection.\arabic{equation}}
}
\begin{document}
\begin{titlepage}

\vspace{-5cm}

\title{
   \hfill{\normalsize UPR-1017-T} \\[1em]
   {\LARGE $SU(4)$ Instantons on  Calabi-Yau Threefolds with ${\mathbb Z}_{2} \times {\mathbb Z}_{2}$ Fundamental Group}
\\
[1em] }
\author{
     Ron Donagi$^2$, Burt A. Ovrut$^1$, Tony Pantev$^2$ and
Ren\'e Reinbacher$^1$ \\[0.5em]
   {\normalsize $^1$Department of Physics, University of Pennsylvania} \\[-0.4em] {\normalsize Philadelphia, PA 19104--6396}\\
   {\normalsize $^2$Department of Mathematics, University of Pennsylvania} \\[-0.4em]
   {\normalsize Philadelphia, PA 19104--6395, USA}\\ }
\date{}
\maketitle
\begin{abstract}
\noindent
Structure group $SU(4)$ gauge vacua of both weakly and strongly
coupled heterotic superstring theory compactified on torus-fibered
Calabi-Yau threefolds $Z$ with ${\mathbb Z}_{2} \times {\mathbb
Z}_{2}$ fundamental group are presented. This is accomplished by
constructing invariant, stable, holomorphic rank four vector bundles
on the simply connected cover of $Z$. Such bundles can
descend either to Hermite-Yang-Mills instantons on $Z$ or to twisted gauge
fields satisfying the Hermite-Yang-Mills equation corrected by
a non-trivial flat $B$-field. It is shown that large families of such
instantons satisfy the constraints imposed by particle physics
phenomenology. The discrete parameter spaces of those families are
presented, as well as a lower bound on the dimension of the continuous
moduli of any such vacuum. In conjunction with ${\mathbb Z}_{2} \times
{\mathbb Z}_{2}$ Wilson lines, these $SU(4)$ gauge vacua can lead to
standard-like models at low energy with an additional $U(1)_{B-L}$
symmetry. This $U(1)_{B-L}$ symmetry is very helpful in naturally
suppressing nucleon decay.

\end{abstract}

\thispagestyle{empty}

\end{titlepage}

\section{Introduction}

The compactification of both weakly coupled \cite{gsw} and strongly
coupled \cite{hw1,hw2,w1} heterotic string theory on smooth Calabi-Yau
threefolds can potentially lead to phenomenologically interesting
particle physics models in the four uncompactified dimensions. In both
cases, the physical properties of the low energy theory are governed
by the $E_{8}$ gauge field vacua on the internal space. These vacua
determine the low energy gauge group and the number of quark and
lepton families. In addition, they contribute moduli which play an
important role in the structure of non-perturbative superpotentials
\cite{bdo1,bdo2} and in vacuum stability \cite{bo}. It follows that
constructing gauge field vacua, that is, instantons with structure
group $G\subseteq E_8$ on Calabi-Yau threefolds, is of fundamental
importance in the attempt to derive the real world from superstring
theory.

It is well known how to construct $G$-instantons on four dimensional
real manifolds and on complex surfaces. This is accomplished via the
so-called ADHM construction \cite{adhm}. This construction yields all
information about the gauge connections and the associated holomorphic
vector bundles. However, until relatively recently, little was known
about how to construct arbitrary $G$-instantons on Calabi-Yau
threefolds. This was due to several factors, not the least of which
was the almost complete lack of knowledge about the metrics on such
manifolds. This situation changed significantly with the work of
\cite{fmw1}, \cite{don}   and \cite{fmw2}  where stable, holomorphic vector
bundles with arbitrary structure group were constructed on simply
connected elliptically fibered Calabi-Yau threefolds. These papers
exploited the theorem of Uhlenbeck and Yau \cite{uyau} and Donaldson
\cite{d} that proves that any such bundle must admit a unique gauge
connection that satisfies the hermitian Yang-Mills equation. That is,
finding $G$-instantons on an elliptically fibered Calabi-Yau threefold
is equivalent to constructing stable holomorphic vector bundles with
structure group $G$. It remains unknown how to solve the hermitian
Yang-Mills equation, but \cite{fmw1}, \cite{don} and \cite{fmw2} open the door
to constructing, relatively easily, the associated vector bundles.

These results were extended and applied to both weakly and strongly
coupled heterotic superstring vacua in a series of papers. In
\cite{low2,dlow,low1}, vector bundles were constructed on simply
connected Calabi-Yau threefolds to produce low energy grand unified
theories with gauge groups such as $E_6$, $SO(10)$ and $SU(5)$.  This
work was carried out within the context of compactified strongly
coupled heterotic string theory, known as heterotic M-theory
\cite{losw1,losw2}. Physically relevant phenomena such as $N=1$
supersymmetry breaking \cite{low4,low3}, five-brane moduli
\cite{dow}, non-perturbative superpotentials \cite{lopr} and small
instanton phase transitions \cite{opp} were explored in these
theories. These concepts were also used to develop a new theory of the
early universe, termed Ekpyrotic cosmology \cite{kost}. Although
originally presented in heterotic M-theory, most of this work applies
to compactifications of the weakly coupled heterotic string as well.

As constructed, these grand unified theories need non-zero vacuum
expectation values to develop in the low energy theory in order to
break to the standard model gauge symmetry.  A more direct way to
produce standard-like models is to allow Wilson lines. Mathematically
this amounts to adding non-trivial flat bundles to the holomorphic vector
bundles.  However, this requires two new objects, torus-fibered
Calabi-Yau threefolds with non-trivial fundamental group and stable,
holomorphic vector bundles on such manifolds, to
be constructed. The first steps in this direction were presented in a
series of papers \cite{dopw-i,dopw-ii,dopw-iii,dopw-iv}. In these
papers, torus-fibered Calabi-Yau threefolds $Z$ with non-trivial
${\mathbb Z}_2$ fundamental group were constructed and relevant
geometrical properties, such as the structure of $H_4(Z,{\mathbb Z})$,
were discussed. Then, a general method for building stable, 
holomorphic vector bundles with structure group $G=SU(5)$ on these
manifolds was presented. The result was standard-like models at low
energy, obtained by breaking the $SU(5)$ low energy gauge group to
$SU(3)_C\times SU(2)_L\times U(1)_Y$ with a ${\mathbb Z}_2$ Wilson
line. These models, although promising, suffer from a potential
deficiency. That is, there is no natural mechanism to suppress nucleon
decay in this context. Although such suppression might be achieved, by
discrete symmetries for example, it has to be implemented on a
restricted, case by case basis.

This has prompted the authors to consider a major generalization of
the work in \cite{dopw-i,dopw-ii,dopw-iii,dopw-iv} to include a much
wider range of fundamental groups and structure groups. The question
of nucleon decay has led us to construct, as the next step,
torus-fibered Calabi-Yau threefolds with ${\mathbb Z}_2\times {\mathbb
Z}_2$ fundamental group supporting stable, holomorphic vector bundles
with $G=SU(4)$ structure group. This choice of 
structure group was motivated by the results of \cite{w2,w3}, where it
was pointed out that such theories could lead to standard-like models
with an additional $U(1)_{B-L}$ gauge group. This $U(1)_{B-L}$
symmetry, although not a complete mechanism for nucleon decay
suppression, is very helpful in suppressing the most egregious
dimension four operators. Also see \cite{rt}. In \cite{opr-i}, we
constructed torus-fibered Calabi-Yau threefolds $Z$ with ${\mathbb
Z}_2\times {\mathbb Z}_2$ and ${\mathbb Z}_2\times {\mathbb Z}_2\times
{\mathbb Z}_2$ fundamental group over a base surface $d{\mathbb
P}_9$. A detailed discussion of their geometry was given and their
moduli spaces presented. An important ingredient in constructing
standard-like models is the homology group $H_4(Z,{\mathbb Z})$. The
explicit construction of this group in the present context is
non-trivial. For this reason, we devoted a second paper \cite{opr-ii}
entirely to a discussion of $H_4(Z,{\mathbb Z})$ and its
properties. However, there remains the fundamental issue of
constructing $G=SU(4)$ gauge instantons on these Calabi-Yau
threefolds. That is explicitly accomplished in this paper.

Specifically, we do the following. In Section~\ref{Z}, we review the
results of \cite{opr-i} and \cite{opr-ii} that will be required in our construction. 
We also use this opportunity to set our  notation. Section~\ref{physics} is devoted
to a general discussion of rank four vector bundles in this
context. First, bundles are constructed on elliptically fibered
Calabi-Yau threefolds $X$ with trivial fundamental group which admit a
freely acting automorphism group ${\mathbb Z}_2\times {\mathbb
Z}_2$. Any such bundle $V$ that is, additionally, equivariant under
${\mathbb Z}_2\times {\mathbb Z}_2$ will then descend to produce a
stable, holomorphic rank four vector bundle $V_Z$ on the torus-fibered
Calabi-Yau threefold $Z=X/({\mathbb Z}_2\times {\mathbb Z}_2)$ with
${\mathbb Z}_2\times {\mathbb Z}_2$ fundamental group. The
Donaldson-Uhlenbeck-Yau  theorem allows us to identify a unique
connection of the bundle $V_{Z}$ as a solution of the hermitian
Yang-Mills equations on $Z$. Equivariance is a very strong constraint
in this context. A weaker condition, which also leads to physical
vacua, is the requirement that $V$ only be invariant under the action
of ${\mathbb Z}_2\times {\mathbb Z}_2$. For a detailed discussion of
invariance versus equivariance see Appendix~\ref{appendix2}. Stable
holomorphic bundles $V$ on $X$ which are invariant, but not
equivariant, give rise to twisted vector bundles $V_{Z}$ on $Z$, where
the twisting is given by a flat $B$-field $\boldsymbol{B}$. See
\cite{c,cks,dp}, Section~\ref{physics} and Appendix~\ref{appendix2}.  
Every such bundle comes equipped with a gauge field $A$ which solves
its string theory equation of motion. In the presence of a
flat $B$-field $\boldsymbol{B}$,   this equation
is a version of the hermitian Yang-Mills equation in which the right
hand side is modified by $\boldsymbol{B}$. That is,  we have $F_{A} =
\boldsymbol{B}\cdot \operatorname{id}$.  See Section~\ref{physics} for
details. Since the $B$-field is flat, we still obtain an  $N=1$
supersymmetric theory in the four dimensional low-energy limit. When
the bundle $V$ is actually equivariant, then the $B$-field is gauge
equivalent to zero and we get back the usual hermitian Yang-Mills
instantons. Mathematically,  the invariant stable  holomorphic bundles on
$X$ correspond to stable,  holomorphic vector bundles on a suitably
defined Deligne-Mumford gerbe over $Z$  \cite{dp},  but we will not
discuss this interpretation here.

We outline the method we will use to produce invariant  bundles, namely, by
extensions from invariant rank two bundles $V_i,\;i=1,2$ on $X$. In
each particular case, when we want to decide whether we are working in
a trivial $B$-field background, we check the resulting bundles for
equivariance as well. The constraints on $V$ required by particle
physics phenomenology, such as three families of quarks and leptons,
are presented and discussed. It was shown, for example in
\cite{schoen}, that any threefold $X$ of the above structure must be a
fiber product $X=B\times_{\cp{1}}B^{'}$ of two rational elliptic
surfaces $B$ and $B^{'}$. Each invariant rank two bundle $V_i$ on $X$
is explicitly composed from a rank two invariant bundle $W_i$ on $B$
and an invariant line bundle $L_i$ on $B^{'}$. The bundles $W_i$ and
$L_i$ are constructed, their moduli spaces described, and their Chern
character calculated in Section~\ref{Wbundles} and
Section~\ref{lbundles} respectively. Using this data, in
Section~\ref{Vbundles} rank two invariant vector bundles $V_i$ are
produced on $X$. These bundles are then used to construct ${\mathbb
Z}_2\times {\mathbb Z}_2$ invariant rank four vector bundles $V$ on
$X$ by extension. In order for such vector bundles to admit a
connection satisfying the hermitian Yang-Mills condition, it is
necessary that they have the property of stability. The concept of
stable holomorphic vector bundles is recalled in Section~\ref{stable}
and a necessary condition for stability presented. A more complete,
technical description of stability is given in
Appendix~\ref{appendix1}. Section~\ref{numerics} is devoted to
explicitly demonstrating that there is a large class of rank four
vector bundles $V$ on $X$ that satisfy the standard model constraints
and the necessary condition for stability. The discrete parameter
space of these vector bundles is presented. In Section~\ref{Stable} we
prove that, subject to a mild restriction on this parameter space, all
of these ${\mathbb Z}_2\times {\mathbb Z}_2$ invariant holomorphic
vector bundles are stable. This section relies on the technical
discussion of stability given in Appendix~\ref{appendix1}. Finally, in
Section~\ref{conclusions} we draw all of these results together into a
conclusion. We present a large class of rank four, stable holomorphic
vector bundles $V$ on $X$ that are ${\mathbb Z}_2\times {\mathbb Z}_2$
invariant and explicitly give their discrete parameter space. We also
use results derived throughout the paper to put a lower bound on the
complex dimension of the continuous moduli space of any such bundle
$V$ and give a concrete example. As we have already explained, any such
bundle $V$ descends to a twisted hermitian Yang-Mills connection of
standard model type on the torus-fibered Calabi-Yau threefold
$Z=X/({\mathbb Z}_2\times {\mathbb Z}_2)$ with ${\mathbb Z}_2\times
{\mathbb Z}_2$ fundamental group, thus accomplishing our goal. In the
process,  we also analyze  the conditions for the standard
model vacuum to appear in a trivial $B$-field background. We find that,
in our setting, the triviality of the $B$-field is incompatible with the
three generation condition. Although it is likely that our method can
be modified  to produce standard model bundles with trivial $B$-field
background,  we will not pursue this in the present paper since the
phenomenologically relevant constraints coming from the standard
model are insensitive to this issue.

Admittedly, this paper is rather technical. This arises from the fact
that one is, in principle, solving the very non-linear hermitian
Yang-Mills equation on complicated Calabi-Yau threefolds whose metrics
are unknown. From this point of view, it seems remarkable that this
can be accomplished at all, let alone with the relatively
straightforward mathematical techniques used in this paper. Be that as
it may, the results of this paper are immediately relevant to pure
particle physics, opening the door to the construction of
standard-like models with suppressed nucleon decay within the context
of heterotic superstring theory. In several upcoming papers, we  extend
our results to more general fundamental groups and structure groups and 
give a detailed analysis of the Wilson line symmetry breaking patterns
\cite{dlor}. We will also  present standard-like models with $U(1)_{B-L}$
symmetry \cite{dopr-iii} within this context.

\section{Calabi-Yau Threefolds $Z$ with ${\mathbb Z}_2 \times {\mathbb
Z}_2$  Fundamental Group}\label{Z}

Calabi-Yau threefolds $Z$ with non-trivial first homotopy group
\be\label{2.1} \pi_1(Z)={\mathbb Z}_2 \times {\mathbb Z}_2 \ee can be
constructed as follows. Let $X$ be a Calabi-Yau threefold with trivial
fundamental group that admits a freely acting ${\mathbb Z}_2 \times
{\mathbb Z}_2$ group of automorphisms preserving the volume form. Then
the quotient \be Z=X/({\mathbb Z}_2 \times {\mathbb Z}_2) \ee is a
smooth Calabi-Yau threefold with first homotopy group
(\ref{2.1}). There are many types of Calabi-Yau threefolds $X$ with
this property. However, for the construction of standard-like model
vacua in heterotic superstring theory, it is expedient \cite{dlow} to
choose $X$ to be elliptically fibered. That is, $X$ is fibered over a
base $B^{'}$, \be \pi: X \to B^{'}, \ee with the generic fiber of
$\pi$ being isomorphic to a smooth torus $T^2$. Furthermore, this
fibration admits a global section \be \sigma: B^{'}\to X, \ee which
turns each generic fiber into an elliptic curve. The mapping $\sigma $
is referred to as the zero section of $X$. As discussed in
\cite{opr-i}, choosing $X$ to be elliptically fibered renders
$Z=X/({\mathbb Z}_2 \times {\mathbb Z}_2)$ to be torus fibered, since
$Z$ does not admit a global section.

The base space $B^{'}$ of an  elliptically fibered Calabi-Yau threefold is restricted to be either a del Pezzo surface, $d{\mathbb P}_r$ for $r=1,\dots,8$, a Hirzebruch surface $F_r$ for any non-negative integer $r$, an Enriques surface $E$, certain blow-ups of these surfaces or a rational elliptic surface  $d{\mathbb P}_9$. However, for the purpose of constructing Calabi-Yau threefolds admitting freely acting automorphism groups with ${\mathbb Z}_2$  factors, it is particularly convenient \cite{opr-i}  to choose
\be\label{2.5}
B^{'}\cong d{\mathbb P}_9.
\ee
The reason for this is the following. Recall that any  rational elliptic surface is itself elliptically fibered over a $\cp{1}$ base with the  projection map 
\be
\beta^{'}: B^{'}\to {\mathbb P}^{1'}.
\ee
Using this fact, it was shown, for example  in  \cite{opr-i}, that any Calabi-Yau threefold $X$  elliptically fibered over  base (\ref{2.5}) must be the  fiber product over $\cp{1}$ of two rational elliptic surfaces $B$ and $B^{'}$. That is,
\be\label{2.7}
X=B\times_{\cp{1}}B^{'}.
\ee
The fiber product permits the natural projections 
\be
\xymatrix{
& X \ar[dl]_-{\pi'} \ar[dr]^-{\pi} & & \\
B \ar[dr]_-{\beta} & & B' \ar[dl]^-{\beta'}\ar@{}[r]|-{.} & \\
& \cp{1} & &
}
\ee
For any rational elliptic surface $B$, it is relatively straightforward to construct all of the involutions
\be
\tau_B: B \to B,
\ee
where $\tau_B^{2}=id$. This was carried out in \cite{opr-i}. By appropriately restricting the moduli space of $B$, one can find surfaces that admit an  involution and, hence, a ${\mathbb Z}_2 $ automorphism group. Further restrictions of the moduli space allow several commuting involutions, leading to automorphism groups  ${\mathbb Z}_2 \times {\mathbb Z}_2$, ${\mathbb Z}_2 \times {\mathbb Z}_2\times {\mathbb Z}_2$ and so on. With this in mind, one chooses two rational elliptic surfaces $B$ and $B^{'}$ each, for example, admitting a ${\mathbb Z}_2 \times {\mathbb Z}_2$ automorphism group. Let $\tau_{Bi}$ and $\tau_{B^{'}i}$ for $i=1,2$ be the associated generators. Then, as shown in \cite{opr-i}, the Calabi-Yau threefold $X$ constructed as the fiber product (\ref{2.7}) of $B$ and $B^{'}$ inherits a freely acting ${\mathbb Z}_2 \times {\mathbb Z}_2$ automorphism group generated by
\be\label{invX}
\tau_{Xi}=\tau_{Bi}\times_{\cp{1}}\tau_{B^{'}i},\;i=1,2,
\ee
as desired. This explains our choice of $B^{'}\cong d{\mathbb P}_9$ as the base surface of $X$.

The explicit choice of the rational elliptic surfaces $B$ and $B^{'}$ required to produce standard-like model vacua in heterotic superstring theory is rather subtle. In this paper, we are interested in  producing realistic theories with ${\mathbb Z}_2 \times {\mathbb Z}_2$ Wilson lines. Therefore, we must choose $B$ and $B^{'}$ to each admit an automorphism group  containing ${\mathbb Z}_2 \times {\mathbb Z}_2$. As discussed in \cite{opr-i}, there exists a three parameter family of rational elliptic surfaces $B$ that admit exactly ${\mathbb Z}_2 \times {\mathbb Z}_2$ automorphism groups. However, as will become clear later in this paper, such surfaces do not generically lead to standard-like model vacua. To achieve such vacua, it is essential to further restrict $B$ and $B^{'}$ to each be contained in a specific  two-parameter sub-family of rational elliptic  surfaces. This two parameter sub-family was described in detail in \cite{opr-i}. Here, we will simply recall the relevant properties. Let $B$ be a $d{\mathbb P}_9$ surface in this two parameter sub-family. Then $B$ necessarily has two properties shared by all rational elliptic surfaces. First, $B$ is an elliptic fibration
\be\label{2.11}
\beta: B \to \cp{1}.
\ee
Second, the second homology group is given by
\be
H_2(B,{\mathbb Z})\cong {\mathbb Z}l \oplus \bigoplus_{i=1}^{9}e_i,
\ee
where $l$ is the pull-back of a line in $\cp{2}$ and $e_i,\;i=1,\dots,9$ are a set of divisors of $B$ with self-intersection $-1$. These divisors each intersect every fiber of (\ref{2.11}) only once. Note, however, that for $B$ in the two parameter sub-family, not all of the $e_i$ are sections of this fibration. More precisely, $e_3, e_5$ and $e_8$ each consist of two components.  Hence, they are not irreducible and smooth, properties required of any section.

Generic $B\cong d{\mathbb P}_9 $ have twelve singular $I_1$ fibers. However, as shown in \cite{opr-i},  the restriction of $B$ to the two parameter sub-family coalesces these fibers pairwise to produce six $I_2$ Kodaira fibers. These $I_2$ fibers are reducible, each being  of the form
\be
n_i \cup o_i
\ee
for $i=1,\dots,6$. Another important consequence of restricting $B$ to the two parameter sub-family is that the maximal number of commuting involutions is increased, from two in the three parameter case, to three. That is, $B$ has a  ${\mathbb Z}_2 \times {\mathbb Z}_2\times {\mathbb Z}_2$ automorphism group generated by $\tau_{Bi},\;i=1,2,3$. Be that as it may, only two generators,
\be\label{2.14}
\tau_{Bi},\;i=1,2,
\ee
lift to a freely acting involution on $X$. Hence, we consider only the ${\mathbb Z}_2 \times {\mathbb Z}_2$ sub-group of automorphisms generated by (\ref{2.14}). It was shown in \cite{opr-i} that these generators have the form
\be\label{ex042}
\tau_{B1}=t_{e_6}\circ \alpha_B,\;\;\;\tau_{B2}=t_{e_4}\circ \alpha_B
\ee
where $\alpha_B$ is an involution on $B$ that leaves the zero section $e_9\equiv e$ invariant and $e_4, e_6$ are sections of $B$ that intersect each fiber at a point of order two. Note that
\be
\tau_{B1}\circ\tau_{B2}=t_{e_4+e_6}
\ee
which acts as a pure translation on the fibers.
%Since only a ${\mathbb Z}_2\times {\mathbb Z}_2$ sub-group of automorphisms li%fts to a free action on $X$, what, then, is the purpose of restricting $B$ to %the two parameter sub-family? The reason is the following.

A class in $H_2(B,\mathbb Z)$ is called invariant if it is not transformed under the action of the ${\mathbb Z}_2 \times {\mathbb Z}_2$ automorphism group. Denote by
\be
H_2(B,\mathbb Z)^{inv} \subset H_2(B,\mathbb Z)
\ee
the homology sub-space of invariant classes. It was shown in \cite{opr-ii} that for $B$ in the two parameter sub-family 
\be
\rank H_2(B,\mathbb Z)^{inv}=4.
\ee
Furthermore, a set of generators of $H_2(B,\mathbb Z)^{inv}\otimes {\mathbb Q}$ is given by
\be
i,\;f,\;n_1+o_2,\;M,
\ee
where 
\be
\begin{split}
i=&2e_6+2e_4-n_1+o_2,\\
M=&2e_2-2e_9-n_1-n_2,\\
\end{split}
\ee
$f$ is the fiber class of $\beta: B \to \cp{1}$ and $n_1, n_2$ and $o_2$  are  components of some $I_2$ fibers.
This choice of generators is motivated by their intersection numbers displayed in Table~\ref{1.6}. Note, in particular, that $M$ is orthogonal to the three other classes. 
\begin{table}[!ht]
\begin{center}
\begin{tabular}{|c||c|c|c|c|} \hline
                  & $i$  & $f$ & $n_1+o_2$& $M$  \\ \hline\hline
$i$               & $-4$ & $4$     & $4$              & $0$  \\ \hline
$f$           & $4$  & $0$     & $0$              & $0$  \\ \hline
$n_1+o_2$ & $4$  & $0$     & $-4$             & $0$  \\ \hline
$M$               & $0$  & $0$     & $0$              & $-4$ \\ \hline
\end{tabular}
\caption{The intersection numbers of the four generators of $H_2(B,{\mathbb Z})^{inv}\otimes {\mathbb Q}$.}
\label{1.6}
\end{center} 
\end{table}
It can  easily be seen that these four invariant generators do not generate $H_2(B,{\mathbb Z})^{inv}$. For example, 
\be\label{ex08}
\frac{1}{2}(i+M+f)=e_6+e_4+e_2-e_9-n_1-n_2+f
\ee
is an integral class in $H_2(B,{\mathbb Z})^{inv}$.  We would not have found this class if we had restricted the coefficients of  $i,f, n_1+o_2$ and $M$ to be integers. We will have to deal with issues arising from this subtlety later in this paper.
As discussed above, choosing both $B$ and $B^{'}$ to be in the two parameter sub-family allows one to construct a  Calabi-Yau threefold $X=B\times_{\cp{1}}B^{'}$ that admits a freely acting ${\mathbb Z}_2 \times {\mathbb Z}_2$ group of automorphisms. This is generated by the involutions
\be
\tau_{Xi}=\tau_{Bi}\times_{\cp{1}}\tau_{B^{'}i},\;i=1,2,
\ee
where $\tau_{Bi}$ and $\tau_{B^{'}i}$ for $i=1,2$ are involutions generating ${\mathbb Z}_2 \times {\mathbb Z}_2$ on $B$ and $B^{'}$ respectively. Furthermore, as shown in detail in \cite{opr-ii}, $X$ inherits a homology sub-group 
\be
H_4(X,{\mathbb Z})^{inv} \subset H_4(X,{\mathbb Z}),
\ee
each of whose classes is invariant under ${\mathbb Z}_2 \times {\mathbb Z}_2$ and which has 
\be
\rank H_4(X,{\mathbb Z})^{inv}=7.
\ee
As we will see, both the rank and the structure of $H_4(X,{\mathbb Z})^{inv}$, which derive from the properties of $H_2(B,{\mathbb Z})^{inv}$, are required to achieve standard-like model vacua.
%This explains our restriction of both $B$ and $B^{'}$ to the two parameter sub-family.

\section{Rank Four Vector Bundles on $X$ and $Z$}\label{physics}

Having constructed torus fibered Calabi-Yau threefolds $Z$ with
fundamental group ${\mathbb Z}_2\times {\mathbb Z}_2$, we begin our
discussion of stable, twisted  vector bundles with structure group
$G=SU(4)$ on $Z$. Note, that any rank four holomorphic vector bundle $V_Z$ defined
on $Z=X/({\mathbb Z}_2\times {\mathbb Z}_2)$ must descend from a rank
four vector bundle $V$ on $X$ that is equivariant under the ${\mathbb
Z}_2\times {\mathbb Z}_2$ automorphism group. Therefore, we will
produce vector bundles on $Z$ by constructing equivariant vector
bundles on $X$. More generally, we will consider  twisted hermitian
Yang-Mills vacua on $Z$, with the twisting being given by a flat
$B$-field. Every such vacuum descends from a stable, holomorphic vector
bundle on $X$ which is  ${\mathbb Z}_2\times {\mathbb Z}_2$
invariant,  but not necessarily equivariant. Conversely, every stable
invariant bundle on $X$ descends to a twisted hermitian Yang-Mills
instanton for some flat $B$-field on $Z$. See Appendix~\ref{appendix2}
for a discussion. Geometrically, these vacua correspond to stable
holomorphic bundles on a Deligne-Mumford gerbe on $Z$, which is
determined by the $B$-field. Since the theory of such bundles is rather
technical, we will not discuss it here. However, before  proceeding  with
the construction, it will be instructive to give the general form
of the twisted hermitian Yang-Mills fields as they appear in
physics. Recall that a flat $B$-field in physics is understood as a
locally defined closed two-form which transforms in a specific way
under a group of generalized gauge transformations. Mathematically,
flat $B$-fields are understood as connections on ${\mathbb C}^{*}$ gerbes on
$Z$. These gerbes are parameterized, up to equivalence, by the
cohomology group $H^{2}(Z,{\mathbb C}^{*})$, and the actual $B$-fields are
represented by cocycles for an appropriately chosen model of this
cohomology group. This is analogous to the way a flat
connection on a $U(1)$ bundle is viewed as a cocycle representing
a class in $H^{1}(Z,U(1))$. Furthermore, if $\boldsymbol{B}$ is a flat
$B$-field on $Z$ and $A$ is an  hermitian connection on some
$\boldsymbol{B}$-twisted vector bundle on $Z$, then the modified
hermitian Yang-Mills equation for  $A$ becomes
\be
F_{A} = \boldsymbol{B}\cdot I.
\ee
This is explained in greater detail in
Appendix~\ref{appendix2}.

As mentioned above, it is expedient \cite{dlow} to choose $X$ to be
elliptically fibered. The reason is that on such manifolds there
exists a powerful tool, namely the spectral cover construction
\cite{fmw1,fmw2,don}, for creating large families of holomorphic
vector bundles. Furthermore, straightforward conditions on the
spectral covers ensure stability of the corresponding vector
bundles. It was shown in \cite{fmw2} that a sufficient condition for
stability is simply that the spectral cover be smooth and irreducible.
These conditions were easily fulfilled for vector bundles
corresponding to grand unified theories \cite{low2,dlow}. In these
situations, the vector bundles are not required to be invariant under
a non-trivial automorphism group. In \cite{dopw-i,dopw-ii}, vacua
associated with standard-like models were constructed from rank five
vector bundles. These bundles were required to be equivariant under a
${\mathbb Z}_2$ automorphism group on $X$. It was found that requiring
${\mathbb Z}_2$ equivariance greatly enhances the difficulty of
producing stable, holomorphic bundles using spectral covers. Be that
as it may, a family of ${\mathbb Z}_2$ equivariant stable, holomorphic
vector bundles was so constructed.

In this paper, we want to produce standard-like model rank four,
holomorphic vector bundles $V$ that are stable and invariant under a
${\mathbb Z}_2\times {\mathbb Z}_2$ automorphism group on $X$. Our
initial approach was to attempt to use the spectral cover
construction. However, we found this to be untenable for the following
reason. Recall that when $X$ is elliptically fibered over a base
$B^{'}\cong d{\mathbb P}_9$, it must have the form
$X=B\times_{\cp{1}}B^{'}$. It was shown in \cite{dopw-ii} that, in
this case, the spectral cover construction on $X$ can be reduced to
finding spectral covers on the rational elliptic surface $B$. The
requirement that $V$ be invariant under ${\mathbb Z}_2\times {\mathbb
Z}_2$ puts strong constraints on the allowed spectral covers on
$B$. We find that such spectral covers must be the union of multiples
of the zero section $e: \cp{1}\to B$ and the fiber $f$ of $B$, not
merely homologous to them. But such covers are highly singular and
reducible. Therefore, there is no easy argument for the stability of
the corresponding bundles. Hence, at this point, it seems more
constructive to abandon the spectral cover approach and to build
stable, rank four vector bundles directly. This conclusion would
appear to nullify the reason for choosing $X$ to be elliptically
fibered. However, this is not the case. Although the fiber product
structure $X=B\times_{\cp{1}}B^{'}$ and the requirement that the
bundles be ${\mathbb Z}_2\times {\mathbb Z}_2$ invariant makes it
difficult to construct stable vector bundles via spectral covers, they
greatly enable producing stable vector bundles as non-trivial
extensions, as we now discuss.

In broad outline, we will proceed as follows. Stable rank four vector
bundles $V$ on $X$ will be constructed as non-trivial extensions
\be\label{ex051}
\ses{V_1}{V}{V_2}
\ee
of rank two vector bundles $V_i,\;i=1,2$ on $X$. These are defined in
terms of two rank two bundles $W_i,\;i=1,2$ on $B$ and two line
bundles $L_i,\;i=1,2$ on $B^{'}$. Specifically,
\be\label{ex050}
V_i=\pi^{'*}W_i\otimes \pi^{*}L_i,\;i=1,2.
\ee
Vector bundles $V$ can be ${\mathbb Z}_2\times {\mathbb Z}_2$
invariant even if one, or both, of $V_1$ and $V_2$ are not
invariant. However, bundles $V$ of this type are harder to construct
and we will not discuss them in this paper. On the other hand, if
$V_1$ and $V_2$ are both ${\mathbb Z}_2\times {\mathbb Z}_2$
invariant, then it is relatively straightforward to construct
invariant vector bundles $V$ defined via (\ref{ex051}). Therefore, we
will restrict the discussion in this paper to bundles $V$ created by
extension from ${\mathbb Z}_2\times {\mathbb Z}_2$ invariant bundles
$V_1$ and $V_2$. It follows from the definition of the rank two
bundles $V_1$ and $V_2$ in (\ref{ex050}) that they will be ${\mathbb
Z}_2\times {\mathbb Z}_2$ invariant if the rank two bundles $W_1, W_2$
and the line bundles $L_1,L_2$ are invariant. In subsequent sections,
we will present the construction of invariant bundles $W_i$, $L_i$ and
$V_i$ for $i=1,2$ and explore their properties. Finally, it will be
necessary to prove that non-trivial extensions $V$, created via
(\ref{ex051}) from $V_1$ and $V_2$, exist. This will be done in the
appropriate places.

Before doing this, however, we want to impose certain restrictions on
the (twisted or untwisted) holomorphic vector bundles $V_Z$ and,
hence, $V$ that are required to produce standard-like model
vacua. First, since we are interested in vector bundles with structure
group $G=SU(4)$, we must require that
\be\label{3.8}
c_1(V_Z)=0.
\ee
Second, anomaly cancellation imposes the requirement that
\be\label{3.4}
c_2(TZ)-c_2(V_Z)=[W],
\ee
where $TZ$ is the tangent bundle of $Z$ and $[W]$ represents the class
of some holomorphic curve in $Z$ around which five-branes are wrapped
\cite{dlow}. Note that (\ref{3.4}) coincides with the anomaly
cancellation condition in heterotic M-theory \cite{hw1, hw2} if one
assumes that the $E_8$ vector bundle on the non-observable orbifold
plane is trivial.   When no five branes
are present, condition (\ref{3.4}) reduces to $c_2(V_Z)=c_2(TZ)$,
which is difficult to solve. However, the existence of five-branes
generalizes this to expression (\ref{3.4}). At first view, (\ref{3.4})
would appear to be simply a definition of $[W]$ and, hence, to put no
constraint on $c_2(V_Z)$. However, this is not quite true. As
discussed in
\cite{dlow}, the five-brane class must be effective since it
represents physical five-branes. Therefore, we require that
\be\label{3.6}
c_2(TZ)-c_2(V_Z)\;\; \effective,
\ee
which puts a non-trivial constraint on $c_2(V_Z)$. Finally, in order
for the low-energy theory to contain three families of quarks and
leptons, one must impose \cite{gsw}
\be\label{3.7}
c_3(V_Z)=6.
\ee
Note that in writing these conditions, one needs to interpret the Chern
classes of $V_{Z}$ as cohomology classes on $Z$. In the untwisted case,
$V_{Z}$ is just a regular bundle and  $c_{i}(V_{Z})$ are  the
usual Chern classes which can be computed, for example, as the invariant
polynomials of the field strength of any gauge field on $V_{Z}$. In
the $\boldsymbol{B}$-twisted case, the curvature $F_{A}$ of a twisted
connection $A$ is not a gauge invariant quantity. However, the
expression $F_{A} - \boldsymbol{B}\cdot I$ is gauge invariant and
should be  understood as the appropriate  field strength for  the twisted
gauge field $A$. The invariant polynomials of this
expression define cohomology classes $c_{i}(V_{Z})$ on $Z$. These are  the Chern
classes of the twisted bundle $V_{Z}$. 

As discussed above, the (twisted or untwisted) bundle $V_Z$ on $Z$ will
descend from a stable, rank four vector bundle $V$ on $X$ that is
invariant under the automorphism group ${\mathbb Z}_2\times {\mathbb
Z}_2$. Since $X$ is a cover of $Z$, it follows that $c_i(V)$ is the
pull-back of $c_i(V_Z)$ for $i=1,2,3$ in either case. Thus, on the level
of cohomology classes on $X$, the difference between the twisted  and
the untwisted cases vanishes. Therefore,   from 
(\ref{3.8}), (\ref{3.6}) and (\ref{3.7}) we conclude that stable, rank four
holomorphic vector bundles $V$ on $X$ will lead to standard-like
models only if they obey the topological constraints
\be\label{ex016}
\begin{array}{c l}
{\bf(C1)} & c_1(V)=0,\\
{\bf(C2)} & c_2(TX)-c_2(V)\; \effective,\\
{\bf(C3)} & c_3(V)=24.
\end{array}
\ee
These constraints will be imposed later in this paper.

\section{Invariant Rank Two Vector Bundles on $B$}\label{Wbundles}

In this section, we construct rank two holomorphic vector bundles on
$B$ that are invariant under the ${\mathbb Z}_2\times {\mathbb Z}_2$
automorphism group generated by $\tau_{Bi},\;i=1,2$, and calculate
their Chern characters. More precisely, we will construct the
requisite rank two invariant vector bundles $W_i,\;i=1,2$ as
non-trivial extensions of a rank one torsion free sheaf by a line
bundle. We will find that the most natural invariant rank two bundles
$W$ are also, by construction, equivariant. Therefore, in the
remainder of this paper, all bundles $W$ will be assumed to have both
properties.

To begin, let us consider two line bundles $L_1$ and $L_2$ on
$B$. Their direct sum $L_1\oplus L_2$ is a rank two vector bundle that
fits into a short exact sequence
\be
\ses{L_1}{L_1\oplus L_2}{L_2}.
\ee
The direct sum corresponds to the trivial extension of $L_2$ by
$L_1$. If $\{U_{\alpha}\}$ is an open cover of $B$, then the line
bundles $L_i,\;i=1,2$ can be described in terms of their transition
functions ${g_{\alpha\beta}^i},\;i=1,2$ defined at each intersection
$U_{\alpha}\cap U_{\beta}$. The data for $L_1\oplus L_2$ on an
intersection $U_{\alpha}\cap U_{\beta}$ is given by
\be\label{3.5}
g_{\alpha\beta}^{\oplus}=
\left( \begin{array}{cc}
g_{\alpha\beta}^1 & 0 \\
0   & g_{\alpha\beta}^2\\
\end{array}
\right). 
\ee
Unfortunately, as will be demonstrated later in this paper, rank two
vector bundles of the form $L_1\oplus L_2$ are too constrained to
satisfy the conditions imposed by particle physics
phenomenology. Therefore, we must consider generalizations of
(\ref{3.5}).

If one can modify $g_{\alpha\beta}^{\oplus}$ to
\be\label{0.35}
g_{\alpha\beta}^{\oplus} \to g_{\alpha\beta}^{'}=
\left( \begin{array}{cc}
g_{\alpha\beta}^1 & f_{\alpha\beta}   \\
0   & g_{\alpha\beta}^2 \\
\end{array}
\right), 
\ee
where $g_{\alpha\beta}^{'}$ fulfills the usual  conditions on transition functions, then the new  rank two holomorphic vector bundle $W$ also  fits into  a  short exact sequence of the form
\be\label{2.8}
\ses{L_1}{W}{L_2}.
\ee
Note that the zero in the lower left corner of (\ref{0.35}) is necessary since (\ref{2.8}) implies that $L_1$ is a sub-bundle of $W$. If $f_{\alpha\beta}$ is non-vanishing, then $W$ is called a non-trivial extension of $L_2$ by $L_1$. We denote by
\be
Ext^1_B(L_2,L_1)
\ee
the set of all rank two holomorphic vector bundles $W$ that fit into the exact sequence (\ref{2.8}).  Clearly, $Ext^1_B(L_2,L_1)$ always contains the trivial extension $L_1\oplus L_2$. In principal, $Ext^1_B(L_2,L_1)$ can also contain non-trivial extensions $W$. $Ext^1_B(L_2,L_1)$ can be shown to be a finite dimensional complex vector space. Since each vector space admits an Abelian group structure, $Ext^1_B(L_2,L_1)$ also has a group structure. We need  discuss neither the vector space structure nor the group structure in any detail. Suffices it here to say that the trivial extension $L_1\oplus L_2$ acts as the zero element of the vector space or, equivalently, as the identity element of the Abelian group. 
Note that it is by no means obvious that any non-trivial extensions $W$ exist. To determine if they do, one needs to calculate the vector space $Ext^1_B(L_2,L_1)$ explicitly. To accomplish this, it is useful to note that
\be\label{1.20}
Ext^1_B(L_2,L_1)\cong H^1(B, L_2^{*}\otimes L_1),
\ee
where $L_2^{*}$ denotes the dual of $L_2$. The cohomology group $H^1(B, L_2^{*}\otimes L_1)$ can usually be calculated.

However, as will become clear below, even non-trivial extensions
satisfying (\ref{2.8}) are not sufficient to satisfy the constraints
of particle physics phenomenology. We must go one step further and
consider extensions not of line bundles but, rather, of rank one
torsion free sheaves by line bundles.  Hence, we have to replace the
line bundle $L_2$ in (\ref{2.8}) by a torsion free sheaf. We will give
a complete definition and discuss some properties of torsion free
sheaves in the Appendix~\ref{appendix1}. For now, it is sufficient to
know that every rank one torsion free sheaf on a smooth surface is of
the form
\be\label{ex05}
L\otimes I_{z_k},
\ee
where $L$ denotes a line bundle on the surface and $I_{z_k}$ the ideal sheaf of $k$ points $z_{k}=\{p_1,...,p_k\}$. Note that, in general, the points in  $z_k$ need not be isolated. However, in this paper, we will always restrict ourselves to ideal sheaves $I_{z_k}$ where all of the points of $z_k$ are distinct.
To get an understanding of  ideal sheaves, first recall that the structure sheaf $\oy$ contains, for any open set $U\subset B$, all holomorphic functions $f : U\to \mathbb{C}$. Clearly, $\oy$ is  the trivial holomorphic line bundle on $B$, which explains the notation. 
Now consider the  ideal sheaf $I_p$ of a single point $p$ on $B$. For any open set $U\subset B$ and $p \in U$, $I_p(U)$ contains all holomorphic functions $f : U\to \mathbb{C}$ which vanish on $p$. For any open set $U\subset B$ not containing $p$, $I_p(U)$ contains all holomorphic functions $f : U\to \mathbb{C}$. Hence, $I_p$ is  a sub-sheaf of the structure sheaf $\oy$ of $B$.  Therefore, we can write 
\be\label{2.19}
\ses{I_p}{\oy}{\mathcal{O}_p},
\ee
where $\mathcal{O}_p$ is the quotient sheaf 
\be
\mathcal{O}_p \cong \oy/I_p.
\ee
This quotient sheaf is also called a ``skyscraper'' sheaf supported at  $p$. Note that the fiber of $\mathcal{O}_p$ vanishes everywhere except at $p$, where it is isomorphic to $\mathbb C$. The ideal sheaf $I_{z_k}$ of $k$ distinct points can similarly be defined. Clearly $I_{z_k}$ fits into the short exact sequence
\be
\ses{I_{z_k}}{\oy}{\bigoplus _{i=1}^{k}\mathcal{O}_{p_i}},
\ee
where
\be
\mathcal{O}_{p_i} \cong \oy/I_{p_i}
\ee
for each $p_i$ in $z_k$. 

For any given line bundles $L_1, L_2$ and an ideal sheaf $I_{z_k}$ on $B$, including  the trivial sheaf $\oy$, we now construct $W$ as the extension of $L_2\otimes I_{z_k}$ by $L_1$, that is
\be\label{1.3}
\ses{L_1}{W}{L_2\otimes I_{z_k}}.
\ee
As above, we define 
\be
Ext^1_B(L_2\otimes I_{z_k},L_1)
\ee
to be the set of all extensions fitting into the exact sequence
(\ref{1.3}).  Again, $Ext^1_B(L_2\otimes I_{z_k},L_1)$ is a complex
vector space with the trivial extension $L_1\oplus (L_2\otimes
I_{z_k})$ as the zero element. By construction, an element $W$ of
$Ext^1_B(L_2\otimes I_{z_k},L_1)$ is a sheaf on $B$, but it is not
necessarily a smooth vector bundle. The criterion for $W$ to be a
vector bundle is the following. First, note that a theorem by Serre
\cite{hl} states that every rank two vector bundle on $B$ is an
element of $Ext^1_B(L_2\otimes I_{z_k},L_1)$ for some $L_1,L_2$ and
$I_{z_k}$. $W \in Ext^1_B(L_2\otimes I_{z_k},L_1)$ is a rank two
vector bundle if and only if $W$ is locally free. For the general
definition of locally free sheaves, we refer the reader to the
Appendix~\ref{appendix1}. Here, we will give a criterion for finding
locally free sheaves within the restricted context discussed below.

In this paper, we will not attempt to solve (\ref{1.3}) for arbitrary line bundles $L_1$ and $L_2$. Rather, we will restrict the line bundles  $L_1$ and $L_2$  on $B$ to be of the from 
\be\label{2.21}
L_1=\oy(af),\;\;\;L_2=\oy(bf)
\ee
where $f$ denotes the fiber class of the elliptic fibration $\beta: B \to \cp{1}$ and $a,b \in {\mathbb Z}$. Furthermore, as indicated above, we will always choose the  sheaves $I_{z_k}$ to contain only distinct points. It follows that the extensions $W$ must satisfy
\be\label{ex057}
\ses{\oy(af)}{W}{\oy(bf)\otimes I_{z_k}}
\ee
and form the vector space
\be
Ext^1_B(\oy(bf)\otimes I_{z_k},\oy(af)).
\ee
In addition, we will only be interested in elements $W$ of this vector space that are rank two vector bundles, that is, are locally free. Unfortunately, not every extension of $\oy(bf)\otimes I_{z_k}$ by $\oy(af)$ is locally free. For example, the trivial extension
\be
\oy(af)\oplus (\oy(bf)\otimes I_{z_k})
\ee
is not locally free, as can easily be checked using the definition
given in the Appendix~\ref{appendix1}. Therefore, it is essential that
we ensure that there are elements in $Ext^1_B(\oy(bf)\otimes
I_{z_k},\oy(af))$ which are locally free sheaves. In this context, the
existence of locally free extensions can be established using the
Cayley-Bacharach criterion \cite{hl}. First, this criterion implies
that
\be
\dim_{\mathbb C}Ext^1_B(\oy(bf)\otimes I_{z_k},\oy(af)) \geq 1.
\ee
Now, by definition, the generic elements form a dense, open subset of $Ext^1_B(\oy(bf)\otimes I_{z_k},\oy(af))$. Let $K_B$ denote the canonical bundle of $B$. Then the Cayley-Bacharach criterion states that a  generic element of $Ext^1_B(\oy(bf)\otimes I_{z_k},\oy(af))$ is a locally free sheaf if and only if any global section 
\be\label{2.20}
s \in H^0(B,\oy(af)^{*}\otimes \oy(bf)\otimes K_B )
\ee
which vanishes at $k-1$ points in $z_k$, vanishes at all points. Otherwise, there are no locally free extensions. Note that if there are no sections, or no section vanishes at $k-1$ points, then a generic element is locally free.  Recall from \cite{opr-i} that for rational elliptic surfaces $B$,
\be
K_B \cong \oy(-f).
\ee 
Hence, to analyze  the global sections in (\ref{2.20}) we need to consider $H^0(B,\oy(mf))$ for any integer $m$. Using  $\beta: B \to \cp{1}$,  it follows  from
\be\label{1.0}
\oy(mf)\cong \beta^{*}\ocp(m)
\ee
and from the projection formula that
\be\label{2.18}
H^0(B,\oy(mf))=H^0(\cp{1},\ocp(m)).
\ee
This  implies that all global sections in $H^0(B,\oy(mf))$ are pull-backs of global sections in $H^0(\cp{1},\ocp(m))$. Thus, the  sections in $H^0(B,\oy(mf))$ are constant along any fiber of $\beta: B \to \cp{1}$. Therefore, all  sections which vanish at a single point in a specific fiber, will vanish everywhere along that  fiber. To ensure that  the Cayley-Bacharach condition is satisfied, we will simply choose the points in the  ideal sheaf $I_{z_k}$ so that each fiber of $\beta: B \to \cp{1}$ which contains one of the points   $p_i \in z_{k}$,  contains at least one other such point. Then the global sections in  $H^0(B,\oy(af)^{*}\otimes \oy(bf)\otimes K_B )$ obey the Cayley-Bacharach criterion and we have ensured that a generic element of $Ext^1_B(\oy(bf)\otimes I_{z_k},\oy(af))$ corresponds to a locally free sheaf, that is, to a rank two vector bundle $W$.

Since the rank two vector bundles $W$ defined via (\ref{ex057}) play a vital role in our construction of rank four vector bundles $V$ on $X$, let us be more precise about how many of these rank two vector bundles exist. Given any integers $a,b \in {\mathbb Z}$ and the  ideal sheaf $I_{z_k}$, it is not too hard to calculate the dimension of $Ext^1_B(\oy(bf)\otimes I_{z_k},\oy(af))$. Let us assume that $k$ is even, that is,
\be
k=2r
\ee
for any positive integer $r$. As we  will show below, this choice is a necessary condition for  the invariance of the  rank two bundles under the automorphism group. Furthermore, to assure that the Cayley-Bacharach condition is satisfied, we will distribute 
the $2r$ points  pairwise  over $r$ smooth fibers.  Then, we can use the results of \cite{dopr-ii} where the dimension of  $Ext^1_B(\oy(bf)\otimes I_{z_{2r}},\oy(af))$ is calculated. We summarize the results in Table~\ref{table-dimExt}. Note that these results depend only on
\be
w\equiv a-b \in {\mathbb Z}
\ee
and the positive integer $r$.
\begin{table}
\begin{center}
\begin{tabular}{|c|c|} \hline 
$w $  &  $\dim_{\mathbb C} Ext^{1}_B(\oy(bf)\otimes I_{z_{2r}},\oy(af))$  \\ \hline \hline 
  $> 0$     &  $w+2r$\\
 $ 0$          &  $2r$ \\
$ -1$          & $2r-1+\max{(1-r,0)}$ \\
 $\leq -2$     & $2r-1+\max{(-w-r,0)}$\\ \hline
\end{tabular}
\end{center}
\caption{The dimension of the the vector space $Ext^{1}_B(\oy(bf)\otimes I_{z_{2r}},\oy(af))$ for $a,b \in{\mathbb Z},\;w \equiv a-b$ and $r \in {\mathbb Z}_{>0}$. }\label{table-dimExt}
\end{table}
\noindent
Then, for any fixed integers $r$, $a$ and $b$, the rank two  vector bundles $W$ form a dense, open subset in $Ext^1_B(\oy(bf)\otimes I_{z_{2r}},\oy(af))$.
For concreteness, let us consider two examples which will turn out to be useful later. First assume that
\be
w= -2,\;\;\; r=1.
\ee
Then, it follows from  Table~\ref{table-dimExt} that
\be\label{ex058}
\dim_{\mathbb C} Ext^{1}_B(\oy(bf)\otimes I_{z_{2}},\oy(af))=2.
\ee 
The generic elements in $Ext^{1}_B(\oy(bf)\otimes I_{z_{2}},\oy(af))$ form, by definition, a dense open subset of $Ext^{1}_B(\oy(bf)\otimes I_{z_{2}},\oy(af))$. Therefore, for $w=-2,r=1$ we have a two dimensional moduli space of rank two vector bundles fitting into the short exact sequence (\ref{ex057}).
A second  example is the case
\be
w= -4,\;\;\; r=4.
\ee
It follows from Table~\ref{table-dimExt} that
\be\label{ex059}
\dim_{\mathbb C} Ext^{1}_B(\oy(bf)\otimes I_{z_{8}},\oy(af))=7
\ee 
and, hence, we obtain a seven dimensional moduli space of rank two vector bundles $W$ satisfying (\ref{ex057}).
Having established that there are many rank two vector bundles on $B$,
we now want to find those that are invariant under the ${\mathbb
Z}_2\times {\mathbb Z}_2$ generators $\tau_{B1}$ and $\tau_{B2}$. 
Recall from (\ref{ex057}) that the rank two bundles $W$ are defined
to be extensions of sheaves of the form $\oy(bf)\otimes I_{z_k}$ by
the line bundle $\oy(af)$. It is possible for $W$ to be ${\mathbb
Z}_2\times {\mathbb Z}_2$ invariant even if either, or both, of
$\oy(af)$ and $\oy(bf)\otimes I_{z_k}$ are not invariant.  However,
such bundles $W$ are harder to construct and we will not consider
them. In this paper, we will always choose both $\oy(af)$ and
$\oy(bf)\otimes I_{z_k}$ to be explicitly ${\mathbb Z}_2\times
{\mathbb Z}_2$ invariant. The consequence of this is that there will
be a well-defined ${\mathbb Z}_2\times {\mathbb Z}_2$ action on
$Ext^{1}_B(\oy(bf)\otimes I_{z_{k}},\oy(af))$, as easily seen from the
exact sequence (\ref{ex057}). One can then use this action to
determine the extensions $W$ that are ${\mathbb Z}_2\times {\mathbb
Z}_2$ invariant. Let us begin by giving the conditions for both
$\oy(af)$ and $\oy(bf)\otimes I_{z_k}$ to be ${\mathbb Z}_2\times
{\mathbb Z}_2$ invariant.

The invariance of line bundles of the form $\oy(mf)$ for any integer
$m$ is easy to prove. First, note that the divisor class $f$ is
invariant under $\tau_{B1}$ and $\tau_{B2}$, as shown in
\cite{opr-ii}. This implies that for any $m$ the line bundle $\oy(mf)$ is
invariant and is our motivation for choosing $L_1$ and $L_2$ in
(\ref{2.21}) to be of the form $\oy(mf)$.  In fact, we can prove the
equivariance of these line bundles. Recall from (\ref{1.0}) that
$\oy(mf)\cong \beta^{*}\ocp(m)$. Furthermore,
\be\label{2.22}
\beta\circ \tau_{Bi} = \tau_{\cp{1}} \circ \beta,\;i=1,2,
\ee
where $\tau_{\cp{1}}$ denotes the induced involution on $\cp{1}$ which
has the two fixed points located at $0$ and $\infty$ \cite{opr-i}.
Since every line bundle on ${\mathbb P}^{1}$ is equivariant with
respect to the action of $\operatorname{Aut}({\mathbb P}^{1})$, it
follows that $\ocp(m)$ is $\tau_{\cp{1}}$ equivariant. But, using
(\ref{1.0}) and  (\ref{2.22}) we see that
\be
\tau_{Bi}^{*}\oy(mf)\cong \oy(mf),
\ee
thus proving the equivariance of line bundles of the form $\oy(mf)$
under the involutions $\tau_{Bi},\;i=1,2$.  It remains to check the
equivariance of $I_{z_k}$. The ideal sheaf $I_{z_k}$ will be
equivariant if the set $z_k=\{p_1,\ldots ,p_k\}$ is invariant under
$\tau_{B1}$ and $\tau_{B2}$. To achieve this, we need to make specific
choices of $z_k$. First, consider a generic point $p$ and define $z_4$
to be
\be
z_4=\{p,\tau_{B1}(p),\tau_{B2}(p),\tau_{B1}\circ\tau_{B2}(p)\}.
\ee
By construction, the four-tuples $z_4$ and, hence, the associated sheaves $I_{z_4}$ are equivariant under the action of $\tau_{B1}$ and $\tau_{B2}$. Second, let us choose any  point $q$ satisfying  $q=\tau_{B1}(q)$. There are at least four such points in $B$. To see this, recall from \cite{opr-i} that $\alpha_B$ acts as $\alpha_B(a)=-a$ for any $a\in f_{\infty}$ and, hence, $\alpha_B$ leaves fixed all four points of order two on $f_{\infty}$, including $e_6$. There are four points $q\in f_{\infty}$, with the property that
\be\label{ex040}
2q=e_6.
\ee
Then 
\be
\tau_{B1}(q)=t_{e_6}\circ \alpha_B(q)=e_6-q=q,
\ee
where we have used  (\ref{ex042}) and (\ref{ex040}). Given any such $q$, we define 
\be
z_2=\{q,\tau_{B2}(q)\}.
\ee
By construction, the two-tuples $z_2$ and the associated sheaves
$I_{z_2}$ are necessarily equivariant under $\tau_{B1}$ and
$\tau_{B2}$. Of course, we can also construct invariant two-tuples
$z_2$, and the corresponding equivariant sheaves, by starting with a
point $q^{'}$ that is fixed under $\tau_{B2}$. The sheaves we will
consider in this paper will be generated by a combination of these
methods. For example, choose a general point $p$, some point $q\neq p$
satisfying $q=\tau_{B1}(q)$ and consider the six-tuple
$z_6=\{p,\tau_{B1}(p),\tau_{B2}(p),\tau_{B1}\circ\tau_{B2}(p),q,
\tau_{B2}(q)\}$. Then,
clearly, the associated ideal sheaf $I_{z_6}$ is equivariant under
$\tau_{B1}$ and $\tau_{B2}$. Proceeding in this way, one can construct
ideal sheaves with $m$ four-tuples and $n$ two-tuples, that is
\be\label{ex023}
I_{z_k},\; k=4m+2n, \,m \in {\mathbb Z}_{\geq 0},\;n=0,\dots,4,
\ee
which are equivariant under $\tau_{Bi},\;i=1,2$.  The fact that we
restrict $n$ to be $0\leq n\leq 4$ reflects the following fact. Note
that there are only four points $q^{'}$ on $B$ which are fixed under
$\tau_{B2}$ and lie on a smooth fiber. Since these four points get
pairwise exchanged under $\tau_{B1}$, there are only two different
invariant two-tuples which can be constructed using points invariant
under $\tau_{B2}$. Considering also two-tuples which one obtains
starting with a point invariant under $\tau_{B1}$ explains the bound.
We will restrict the points in $z_k$ to lie on smooth fibers. Then,
for invariant $k$-tuples $z_k$, any fiber containing one point of
$z_k$ contains another such point as well. This follows simply from
the fact that $\tau_{B1}\circ\tau_{B2} $ acts as a translation along
the fibers of $\beta: B \to \cp{1}$. Thus, the Cayley-Bacharach
criterion discussed above is automatically satisfied.  Therefore, the
existence of generic locally free extensions using these equivariant
ideal sheaves $I_{z_k}$ is ensured. Putting everything together, we
conclude the following.

Choose $\oy(af)$ and $\oy(bf)\otimes I_{z_k}$ with 
\be
a,b \in {\mathbb Z},\;k=4m+2n,\;m \in {\mathbb Z}_{\geq 0},\;n=0,\dots,4.
\ee
Then the vector space of extensions $Ext^1_B(\oy(bf)\otimes
I_{z_k},\oy(af))$ contains a dense, open subset of rank two vector
bundles $W$. Since $\oy(af)$ and $\oy(bf)$ are automatically
equivariant under ${\mathbb Z}_2 \times{\mathbb Z}_2 $, and $I_{z_k}$
is chosen to be equivariant under these automorphisms, it follows that
both $\oy(af)$ and $\oy(bf)\otimes I_{z_k}$ are ${\mathbb Z}_2
\times{\mathbb Z}_2 $ equivariant. As a consequence, there is a
natural action of ${\mathbb Z}_2 \times{\mathbb Z}_2 $ on
$Ext^1_B(\oy(bf)\otimes I_{z_k},\oy(af))$. Denote by
\be
Ext^1_B(\oy(bf)\otimes I_{z_k},\oy(af))^{inv}
\ee
the subspace of $Ext^1_B(\oy(bf)\otimes I_{z_k},\oy(af))$ on which
${\mathbb Z}_2 \times{\mathbb Z}_2$ acts trivially.  If such a
subspace is non-trivial, then the ${\mathbb Z}_2\times {\mathbb Z}_2$
equivariant vector bundles $W$ form a dense, open subset of
$Ext^1_B(\oy(bf)\otimes I_{z_k},\oy(af))^{inv}$.

We now proceed to show that if $Ext^1_B(\oy(bf)\otimes
I_{z_k},\oy(af))$ is non-trivial then \linebreak $Ext^1_B(\oy(bf)\otimes
I_{z_k},\oy(af))^{inv}$ not only is non-trivial, but its dimension can
be made to satisfy the constraint
\be\label{ex070}
\dim_{\mathbb C}Ext^1_B(\oy(bf)\otimes I_{z_k},\oy(af))^{inv}\geq \frac{1}{4}\dim_{\mathbb C}Ext^1_B(\oy(bf)\otimes I_{z_k},\oy(af)).
\ee
To show this, first note that the automorphism group ${\mathbb Z}_2\times {\mathbb Z}_2$ acts linearly on the vector space $Ext^1_B(\oy(bf)\otimes I_{z_k},\oy(af))$. Since ${\mathbb Z}_2\times {\mathbb Z}_2$ is Abelian, its irreducible representations are one dimensional and we can diagonalize the action of ${\mathbb Z}_2\times {\mathbb Z}_2$ on $Ext^1_B(\oy(bf)\otimes I_{z_k},\oy(af))$.  There are four different irreducible representations of ${\mathbb Z}_2\times {\mathbb Z}_2$ given by the four characters $\chi_c,\;c=1,\dots,4$. Hence, we have the decomposition
\be
Ext^1_B(\oy(bf)\otimes I_{z_k},\oy(af)) = \oplus_{c=1}^{4}n_c E_c,
\ee
where $E_c$ is the one dimensional eigenspace transforming under the $c-th$ character $\chi_c$ and $n_c$ is its multiplicity. If we choose  the eigenspace $E_{c^{'}}$ with the highest multiplicity $n_{c^{'}}$, then  we can modify the action of ${\mathbb Z}_2\times {\mathbb Z}_2$ on $Ext^1_B(\oy(bf)\otimes I_{z_k},\oy(af))$ so that this eigenspace corresponds to the  trivial representation. It follows that
\be
Ext^1_B(\oy(bf)\otimes I_{z_k},\oy(af))^{inv}\equiv n_{c^{'}}E_{c^{'}}
\ee
is non-trivial and, furthermore, that its dimension is bounded from below as in expression (\ref{ex070}). As stated above, the equivariant vector bundles $W$ form a dense, open subset of $Ext^1_B(\oy(bf)\otimes I_{z_k},\oy(af))^{inv}$.

Again, since these equivariant rank two bundles play a vital role in our construction, let us be even more precise about how many such vector bundles exist. Note that for the sheaves $I_{z_k}$ in (\ref{ex023}), one can always write $k=2r$ for some positive integer $r$. Furthermore, the points of $z_k$ can always be pairwise distributed over $k/2$ smooth fibers. Therefore, we can use Table~\ref{table-dimExt} to calculate the dimension of $Ext^1_B(\oy(bf)\otimes I_{z_k},\oy(af))$. Let us again consider the two examples discussed earlier. In the first example, we chose
\be\label{ex061}
w = -2,\;\;\; r=1
\ee
which implied (\ref{ex058}) that 
\be
\dim_{\mathbb C}Ext^1_B(\oy(bf)\otimes I_{z_2},\oy(af))=2.
\ee
It then follows from (\ref{ex070}) that
\be\label{ex080}
\dim_{\mathbb C}Ext^1_B(\oy(bf)\otimes I_{z_2},\oy(af))^{inv}\geq 1.
\ee
Hence, in this case, we have at least a one dimensional moduli space of equivariant rank two vector bundles $W$. In the second example, 
\be\label{ex062}
w = -4,\;\;\; r=4.
\ee
Then, we found in (\ref{ex059}) that
\be
\dim_{\mathbb C}Ext^1_B(\oy(bf)\otimes I_{z_8},\oy(af))=7
\ee
and, hence,
\be\label{ex081}
\dim_{\mathbb C}Ext^1_B(\oy(bf)\otimes I_{z_8},\oy(af))^{inv}\geq 2.
\ee
That is, in this example, we have at least a three dimensional moduli
space of equivariant rank two vector bundles. And so on. Note that if
we take an extension corresponding to a non-trivial character of
${\mathbb Z}_{2}\times {\mathbb Z}_{2}$, then the corresponding vector
bundle $W$ will be automatically invariant but  not necessarily
equivariant. Of course, the dimension estimate above remains 
valid regardless of whether we work with invariant or equivariant
bundles.

Having constructed equivariant rank two vector bundles $W$ on $B$, we
now proceed to calculate their Chern characters. Since the $W$ fit
into the short exact sequence
\be
\ses{\oy(af)}{W}{\oy(bf)\otimes I_{z_k}},
\ee
it follows that their  Chern characters are given by
\be\label{1.4}
\ch W=\ch\oy(af)+\ch\oy(bf)\cdot \ch I_{z_k}.
\ee
The definition of the Chern character of vector bundles and its
interpretation in term of gauge invariant polynomials is standard and
can be found, for example, in \cite{n}. We note that for any line
bundle $L$ on a surface,
\be
\ch L =1+c_1(L)+\frac{1}{2}c_1(L)^2.
\ee
Since $c_1(\oy(mf))=mf$ and $f^2=0$, we find for the Chern character of $\oy(mf)$ that
\be\label{1.5}
\ch \oy(mf)=1+mf.
\ee
However, the ideal sheaf $I_{z_k}$ is not locally free and, hence, is
not a vector bundle. Instead, it belongs to a larger class of sheaves
on $B$, called coherent sheaves. How does one define the Chern
character on coherent sheaves? The relevant point in this regard is
that every coherent sheaf $F$ on a smooth algebraic variety has a
finite resolution by locally free sheaves. That is, there exists an
exact sequence
\be
0\to V_n \to V_{n-1}\to \ldots \to V_0\to F \to 0 
\ee
for some finite integer $n$ such that all $V_i, i=0,\dots,n$ are
vector bundles. One then defines $\ch F$ as
\be
\ch F=\sum_{i=0}^n (-1)^{i}\ch V_i,
\ee
where $\ch V_i$ can be calculated using, for example, a connection on
$V_i$. It can be shown that $\ch F$ does not depend on the chosen
resolution and, hence, is well-defined. This gives a definition for
$\ch I_{z_k}$. To actually calculate $\ch I_{z_k}$, however, it is
more convenient to use the Grothendieck-Riemann-Roch theorem
\cite{f}. This theorem states the following. Let $X$ and $Y$ be smooth
algebraic varieties and $h: X \to Y$ a smooth, proper mapping between
them with the property that $h^{-1}(y)$ contains only a finite number
of points for every $y \in Y$. Then, for any coherent sheaf $F$ on $X$
there is a relation between the Chern character $\ch F$ on $X$ and the
Chern character of its push forward $h_{*}F$ on $Y$. This relation is
given by
\be\label{2.24}
\ch h_{*}F=h_{*}(\ch F\cdot \td X)\cdot \td^{-1}Y.
\ee
Here $\td X $ and $\td Y $ denote the Todd classes of $X$ and $Y$
respectively. Loosely speaking, the Grothendieck-Riemann-Roch theorem
fixes the ``non-commutativity'' of $\ch\cdot h_{*}$ and $h_{*}\cdot
\ch$. Let us first apply this theorem to the calculation of the Chern
character of an ideal sheaf $I_p$ of a point. Recall from (\ref{2.19})
that $I_p$ fits into the short exact sequence
\be
\ses{I_p}{\oy}{i_{*}\mathcal{O}_p}, 
\ee
where $i: p \to B$ is the inclusion.
Hence,  its Chern character is given by
\be\label{2.27}
\ch I_p=\ch\oy-\ch i_{*}\mathcal{O}_p.
\ee
One can now use the Grothendieck-Riemann-Roch theorem to calculate
$\ch i_{*}\mathcal{O}_p$. To do this, take $X=p$, $Y=B$ and $h$ to be
the inclusion map
\be\label{2.25}
i: p \to B.
\ee
Therefore, using (\ref{2.24}) and (\ref{2.25}) we find
\be\label{2.26}
\ch\mathcal{O}_p=\ch i_{*}{\mathcal{O}_p}=i_{*}(\ch{\mathcal{O}}_p\cdot \td p)\cdot \td^{-1}B.
\ee
Since ${\mathcal{O}}_p$ is the trivial bundle on $p$, its Chern
character is simply unity. Similarly, since the tangent bundle of a
point is trivial, we obtain
\be
\td(p)=1.
\ee
The inclusion $i_{*}(1)$ corresponds to a point in $B$, namely, to the
image $i(p)$. The intersection of $i(p)$ with $\td^{-1} B$ vanishes
for all classes in $\td^{-1} B$ except for $1$.  This intersection
clearly corresponds to a point in $B$. Therefore, it follows from
these results and (\ref{2.26}) that
\be
\ch\mathcal{O}_p= \pt.
\ee
Using this result, and the fact that 
\be
\ch \oy =1,
\ee
it follows from (\ref{2.27}) that
\be
\ch I_p =1-\pt.
\ee
For an ideal sheaf $I_{z_k}$ with $k$ distinct points, this result easily generalizes to
\be
\ch I_{z_k}=1-k \pt.
\ee
Using this, (\ref{1.4}) and (\ref{1.5}), we find that the Chern
character of $W$ is given by
\be\label{2.30}
\begin{split}
\ch W  &=1+af+(1+bf)(1-k\pt)\\
       &=2+(a+b)f-k\pt.
\end{split}
\ee

To conclude, to construct a rank four bundle $V$ on $X$ we need to
produce two rank two bundles on $B$. We denote these as $W_i,\;i=1,2$,
and define them to be locally free extensions satisfying
\be
\ses{\oy(a_if)}{W_i}{\oy(b_if)\otimes I_{z_{k_i}}},\;i=1,2
\ee
where $a_i, b_i \in {\mathbb Z}$. If, in addition, the ideal sheaves
are chosen so that $k_i=4m_i+2n_i$ where $m_i \in {\mathbb Z}_{\geq
0},\;n_i=0,\dots,8$ then, by the above discussion, for each $i=1,2$
there exists a subspace $Ext^1_B(\oy(bf)\otimes
I_{z_k},\oy(af))^{inv}$ of $Ext^1_B(\oy(bf)\otimes I_{z_k},\oy(af))$
of complex dimension at least one on which ${\mathbb Z}_2 \times
{\mathbb Z}_2$ acts trivially. Any generic element of
$Ext^1_B(\oy(bf)\otimes I_{z_k},\oy(af))^{inv}$ is a ${\mathbb Z}_2
\times {\mathbb Z}_2$ equivariant rank two vector bundle on $B$. The
associated Chern characters are given by
\be
\ch W_i=2+(a_i+b_i)f-k_i\pt,\;i=1,2.
\ee
In the next section, we proceed  to calculate equivariant line bundles on the surface $B^{'}$.

\section{Invariant Line Bundles on $B^{'}$}\label{lbundles}

Denote by $Pic(B^{'})$ the space of all line bundles on the surface
$B^{'}$ up to isomorphism. Consider the map
\be\label{3.0}
c_1: Pic(B^{'}) \to H^2(B^{'},\mathbb Z),
\ee
where $c_1(L)$ is the first Chern class of any $L \in
Pic(B^{'})$. Since $B^{'}$ is simply connected, the mapping
(\ref{3.0}) is injective. Furthermore, since $B^{'}$ is a rational
elliptic surface this map is also surjective. Therefore, $c_1$
determines an isomorphism
\be\label{3.2}
Pic(B^{'})\cong H^2(B^{'},{\mathbb Z}).
\ee
Using Poincare duality, we know that $H^2(B^{'},{\mathbb Z}) \cong
H_2(B^{'},{\mathbb Z})$ and, hence,
\be\label{3.1}
Pic(B^{'})\cong H_2(B^{'},{\mathbb Z}).
\ee
It follows that every class in $H_2(B^{'},{\mathbb Z})$ determines a
line bundle on $B^{'}$ up to isomorphism. If we denote by $c_1(L)$
both the first Chern class of $L$ and its Poincare dual, then
(\ref{3.1}) implies that any line bundle $L \in Pic(B^{'})$ is
uniquely determined by $c_1(L) \in H_2(B^{'},{\mathbb Z})$.
Therefore, to describe all line bundles on $B^{'}$, we must find an
integral basis of $H_2(B^{'},{\mathbb Z})$. Such a basis was given in
Section~\ref{Z}, where we established that
\be\label{1.7}
H_2(B^{'},{\mathbb Z})\cong {\mathbb Z}l^{'} \oplus \bigoplus_{i=1}^{9}e_i^{'},
\ee
with $l^{'}$ being the pull-back of a line in $\cp{2}$ and
$e_i^{'},\;i=1,\dots,9$ being divisors on $B^{'}$ with
self-intersection $-1$ which intersect each fiber of $\beta^{'}: B^{'}
\to \cp{1}$ only once.

Therefore finding all line bundles on $B^{'}$ which are invariant
under $\tau_{B^{'}1}$ and $\tau_{B^{'}2}$, is equivalent to finding
all classes in $H_2(B^{'},{\mathbb Z})$ which are invariant under
these involutions. The set of all invariant classes is denoted by
$H_2(B^{'},{\mathbb Z})^{inv} \subset H_2(B^{'},{\mathbb Z})$. As
discussed in Section~\ref{Z}, for $B^{'}$ in the two parameter
sub-family,
\be
\rank H_2(B^{'},{\mathbb Z})^{inv} =4.
\ee
Furthermore, a set of four generators of $H_2(B^{'},{\mathbb Z})^{inv}
\otimes {\mathbb Q}$ is given by
\be
i^{'},\;f^{'},\;n^{'}_1+o^{'}_2,\;M^{'},
\ee
where
\be\label{b.6}
\begin{split}
i^{'}=&2e^{'}_6+2e^{'}_4-n^{'}_1+o^{'}_2,\\
M^{'}=&2(e^{'}_2-e^{'}_9)-n^{'}_1-n^{'}_2,\\
\end{split}
\ee
$f^{'}$ denotes the fiber class of $\beta^{'}: B \to \cp{1}$ and
$n^{'}_1, n^{'}_2$ and $o^{'}_2$ are components of some $I_2$ fibers.
Since these generators span $H_2(B^{'},{\mathbb Z})^{inv} \otimes
{\mathbb Q}$, it follows that the most general form of a class in
$H_2(B^{'},{\mathbb Z})^{inv}$ is given by
\be\label{1.8}
\tilde{a}i^{'}+xf^{'}+y(n_1^{'}+o^{'}_2)+\bar{a}M^{'}
\ee
for $\tilde{a},x,y,\bar{a}\in \mathbb{Q}$ such that (\ref{1.8})
expanded in the basis (\ref{1.7}) has integral coefficients.  It
follows from (\ref{3.2}), (\ref{3.1}) and (\ref{1.8}) that $L \in
Pic(B^{'})$ is invariant under $\tau_{B^{'}i},i=1,2$ if and only if
$c_1(L) \in H_2(B^{'},{\mathbb Z})^{inv}$, that is,
\be
c_1(L)=\tilde{a}i^{'}+xf^{'}+y(n_1^{'}+o^{'}_2)+\bar{a}M^{'}.
\ee
As explained above, this expression uniquely determines the line
bundle $L$. The Chern character of $L$ can be written as
\be
\ch L=1_{B^{'}}+c_1(L)+\frac{1}{2}c_1(L)^2.
\ee
In fact, it is easy to show  that the line bundles $\oz(f')$,
$\oz(n_1^{'}+o^{'}_2)$, $\oz(i')$ and $\oz(M')$ are actually
${\mathbb Z}_{2}\times {\mathbb Z}_{2}$ equivariant. Indeed, we
already checked this for $\oz(f')$. To show  the equivariance of 
$\oz(n_1^{'}+o^{'}_2)$, we  need only notice that the effective
divisor $n_1^{'}+o^{'}_2$ is a pull-back of a curve in the quotient
surface $B'/({\mathbb Z}_{2}\times {\mathbb Z}_{2})$. Finally, for the
equivariance of $\oz(i')$ and $\oz(M')$ we observe that,  modulo $f'$ and  $n_1^{'}+o^{'}_2$,  the divisors $i'$ and $M'$ are
divisible by two. Thus, modulo equivariant line bundles, the 
bundles $\oz(i')$ and $\oz(M')$ are tensor squares of invariant line
bundles. Since the obstruction for a  ${\mathbb Z}_{2}\times
{\mathbb Z}_{2}$ invariant bundle to be
equivariant is classified by an element in ${\mathbb Z}_{2}$, see
Appendix~\ref{appendix2}, this  implies   the desired equivariance of 
$\oz(i')$ and $\oz(M')$.

As discussed previously, to construct rank four vector bundles $V$ on
$X$, we will need to choose two invariant line bundles $L_i,\;i=1,2$
on $B^{'}$. We do this by specifying
\be\label{2.31}
c_1(L_i)=\tilde{a}_ii^{'}+x_if^{'}+y_i(n_1^{'}+o^{'}_2)+\bar{a}_iM^{'},\;i=1,2
\ee
with $\tilde{a}_i,x_i,y_i,\bar{a}_i \in {\mathbb Q}$ chosen so that
$c_1(L_i),\;i=1,2$ are integral classes.
It is important to note that, even through the bundles $\oz(i')$, $\oz(f')$,  $\oz(n_1^{'}+o^{'}_2)$ and $\oz(M')$ are individually equivariant, this is not necessarily a property of the linear combination (\ref{2.31}). A line bundle $L_i$ will be equivariant if each of the coefficients $\tilde{a}_i,x_i,y_i,\bar{a}_i$ are integers. However, if any of these coefficients is rational then $L_i$, although invariant by construction, need not be equivariant. In fact, the standard-like model conditions will, in this paper, require that $L_i$ be invariant, but not equivariant.

\section{Invariant Rank Four Vector Bundles on $X$}\label{Vbundles}

In this section, we will construct rank four holomorphic vector
bundles $V$ on $X$ that are invariant under the ${\mathbb Z}_2\times
{\mathbb Z}_2$ automorphism group generated by $\tau_{Xi},i=1,2$ and
calculate their Chern character. We will do this using a method
similar to that employed to construct the rank two vector bundles $W$
on $B$ in Section~\ref{Wbundles}. That is, we will construct $V$ by
extension.

To begin with, let us consider any pair of rank two holomorphic vector
bundles $V_1$ and $V_2$ on $X$. Then, we define a rank four
holomorphic vector bundle $V$ by the exact sequence
\be\label{ex071}
\ses{V_1}{V}{V_2}.
\ee
Denote by
\be
Ext^{1}_X(V_2,V_1)
\ee
the set of all rank four holomorphic vector bundles $V$ that fit into
the exact sequence (\ref{ex071}). $Ext^{1}_X(V_2,V_1)$ always contains
the trivial extension $V_1\oplus V_2$. In principle,
$Ext^{1}_X(V_2,V_1)$ can also contain non-trivial extensions $V$. As
discussed previously, $Ext^{1}_X(V_2,V_1)$ can be shown to be a finite
dimensional vector space. In particular, it has the structure of an
Abelian group. The trivial extension $V_1\oplus V_2$ acts as the zero
element of the vector space or, equivalently, as the identity element
of the Abelian group. Note that it is by no means obvious that any
non-trivial extensions $V$ exists. To determine if they do, one needs
to calculate the vector space $Ext^{1}_X(V_2,V_1)$ explicitly. The
techniques required for doing this go beyond those considered in this
paper. Suffices it here to say that for the rank two bundles $V_1$ and
$V_2$ that we will consider in this paper, we find that
\be\label{ex072}
\dim_{\mathbb C}Ext^{1}_X(V_2,V_1)\geq 1.
\ee
That is, the sequence (\ref{ex071}) always has non-trivial
extensions. This is proven in \cite{dopr-ii}.  Note that in
Section~\ref{Wbundles}, the rank two extensions $W$ were torsion free
sheaves. Only a dense, open subset of these were, additionally,
locally free and, hence, smooth vector bundles. Here, we do not have
this problem. All elements of $Ext^{1}_X(V_2,V_1)$ are holomorphic
vector bundles.

Having shown that there are many rank four holomorphic vector bundles
on $X$, we now want to find those that are invariant  and, possibly, 
equivariant  under the ${\mathbb Z}_2\times {\mathbb Z}_2$ generators
$\tau_{X1}$ and $\tau_{X2}$. Henceforth, we will choose $V_1$ and
$V_{2}$ to be either simultaneously invariant, but not
equivariant, or simultaneously ${\mathbb Z}_2\times {\mathbb
Z}_2$ equivariant. From the defining sequence (\ref{ex071}) for $V$ and the
fact that the obstruction to ${\mathbb Z}_2\times {\mathbb Z}_2$
equivariance is an element in ${\mathbb Z}_{2}$, see
Appendix~\ref{appendix2}, it follows that there is a group action of
${\mathbb Z}_2\times {\mathbb Z}_2$ on $Ext^{1}_X(V_2,V_1)$. As before,
the space $Ext^{1}_X(V_2,V_1)$ will split into eigenspaces labeled by
the characters of ${\mathbb Z}_2\times {\mathbb Z}_2$ and every
element in such an eigenspace will give us an extension which is
invariant as a bundle.  The elements in the  subspace
\be
Ext^{1}_X(V_2,V_1)^{inv}
\ee
of $Ext^{1}_X(V_2,V_1)$ which carries the trivial representation of
${\mathbb Z}_2\times {\mathbb Z}_2$ will correspond to invariant
bundles.  By an argument similar to that leading to equation (\ref{ex070}), one can show that 
\be
\dim_{\mathbb C}Ext^{1}_X(V_2,V_1)^{inv}\geq 
\frac{1}{4}\dim_{\mathbb C}Ext^{1}_X(V_2,V_1).
\ee
It follows from (\ref{ex072}) that, for the rank two ${\mathbb
Z}_2\times {\mathbb Z}_2$ invariant bundles $V_1$ and $V_2$ that we
will consider in this paper, there always exists a vector subspace of
rank four, ${\mathbb Z}_2\times {\mathbb Z}_2$ invariant bundles $V$
with
\be
\dim_{\mathbb C}Ext^{1}_X(V_2,V_1)^{inv}\geq 1.
\ee
We now turn to a more explicit construction of these bundles. Again, if 
$V_{1}$ and $V_{2}$ happen, in addition,  to be
equivariant we will get a non trivial equivariant extension $V$.

Specifically, we will use the invariant rank two vector bundles on $B$
and the invariant line bundles on $B^{'}$ to produce rank four
vector bundles $V$ on $X$ which are invariant under the ${\mathbb
Z}_2\times {\mathbb Z}_2$ automorphism group generated by
$\tau_{Xi},\;i=1,2$. We begin by constructing invariant rank two
vector bundles $\mathcal V $ on $X$. Define
\be\label{ex012}
{\mathcal V}= \pi^{'*}W \otimes \pi^{*}L,
\ee
where $W$ is any equivariant rank two vector bundle on $B$ discussed
in Section~\ref{Wbundles} and $L$ is any invariant line bundle on
$B^{'}$ presented in Section~\ref{lbundles}. By construction,
${\mathcal V}$ is a rank two holomorphic vector bundle on $X$. We now
note that it is also invariant under the action
$\tau_{Xi},\;i=1,2$. First, consider the $\pi^{'*}W$ factor. By
assumption, $W$ is invariant under the ${\mathbb Z}_2\times {\mathbb
Z}_2$ automorphism group of the surface $B$. That is
\be\label{ex09}
\tau_{Bi}^{*}W\cong W,\;i=1,2.
\ee
Recall that the bundles $W$ that we use are, in addition, equivariant.
What about the pull-back bundle $\pi^{'*}W$ on $X$? Note that the action of each generator $\tau_{Xi}$ on $X$ induces the action $\tau_{Bi}$ on $B$. That is
\be\label{1.10}
\pi^{'}\circ \tau_{Xi}= \tau_{Bi}\circ \pi^{'}, \;i=1,2.
\ee
This follows directly from the definition of $X$ as the fiber product of two rational elliptic surfaces $B$ and $B^{'}$, and from the definition of the involutions $\tau_{Xi},\;i=1,2$ in terms of involutions $\tau_{Bi}$ and $\tau_{B^{'}i},\;i=1,2$ on $B$ and $B^{'}$ respectively.
Now  define
\be\label{ex015}
f_i=\pi^{'}\circ \tau_{Xi},\;i=1,2.
\ee
Clearly, each satisfies
\be
f_i: X \to B.
\ee
We can use these mappings to pull-back vector bundles from $B$ to $X$. From (\ref{1.10}) and (\ref{ex015}), we can write the pull-back maps $f^{*}_i$ as
\be
f^{*}_i=\tau_{Xi}^{*}\circ\pi^{'*}=\pi^{'*}\circ\tau_{Bi}^{*}, \;i=1,2.
\ee
Applying the maps $f^{*}_i$ to the rank two vector bundle $W$ on $B$, we find
\be\label{1.13}
f^{*}_i W=\tau_{Xi}^{*}(\pi^{'*}W)=\pi^{'*}(\tau_{Bi}^{*}W)=\pi^{'*}W,\;i=1,2.
\ee
Note that, in the last equality, we have used the fact that the rank two bundle $W$ is invariant and, hence, satisfies (\ref{ex09}).
Equating  the second and last terms in (\ref{1.13}),  it follows that 
\be\label{2.33}
\tau_{Xi}^{*}(\pi^{'*}W)=\pi^{'*}W,\;i=1,2.
\ee
Therefore, the invariance of $W$ on $B$ under $\tau_{Bi},\;i=1,2$
ensures the invariance of $\pi^{'*}W$ on $X$ under
$\tau_{Xi},\;i=1,2$. In addition, it is easy to to show that since $W$ is also equivariant, so is $\pi^{'*}W$.

Now consider the $\pi^{*}L$ factor. By assumption, $L$ is invariant
under the ${\mathbb Z}_2\times {\mathbb Z}_2$ automorphism group of
the surface $B^{'}$. That is
\be\label{ex011}
\tau_{B^{'}i}^{*}L\cong L,\;i=1,2.
\ee
What about the pull-back bundle $\pi^{*}L$ on $X$? 
An  argument similar to the above can be applied to show the invariance of $\pi^{*}L$ under $\tau_{Xi},\;i=1,2$. Here, one notes that the action of each $\tau_{Xi}$ on $X$ induces the action of $\tau_{B^{'}i}$ on $B^{'}$, namely
\be
\pi\circ \tau_{Xi}= \tau_{B^{'}i}\circ \pi, \;i=1,2.
\ee
Applying the same reasoning as above and the fact that $L$ is chosen to satisfy (\ref{ex011}), we conclude that
\be\label{2.34}
\tau_{Xi}^{*}(\pi^{*}L)=\pi^{*}L,\;i=1,2.
\ee 
That is, the invariance of $L$ on $B^{'}$ under  $\tau_{B'i},\;i=1,2$ ensures the invariance of $\pi^{*}L$ on $X$ under $\tau_{Xi},\;i=1,2$.

Note that this agrees with our results in  \cite{opr-ii}. There, it was shown that to ensure invariance under $\tau_{Xi},\;i=1,2$ for any divisor on $X$ of the form $\pi^{*}D^{'}$ and $\pi^{'*}D$, where $D$ and $D^{'}$ are divisors on $B$ and $B^{'}$ respectively, we need only to  show the invariance of $D$ and $D^{'}$ under $\tau_{Bi},\;i=1,2$ and $\tau_{B^{'}i},\;i=1,2$ respectively. If one associates, for example, the line bundle $\oz(D^{'})$ to the divisor $D^{'}$ on $B^{'}$, then the line bundle $\pi^{*}\oz(D^{'})$ corresponds to the divisor $\pi^{*}D^{'}$ on $X$. The invariance of $\oz(D^{'})$ and $D^{'}$ is now sufficient to ensure the invariance of $\pi^{*}\oz(D^{'})$ and $\pi^{*}D^{'}$.

Using (\ref{ex012}), the fact that
\be
\tau_{Xi}^{*}{\mathcal V}\cong \tau_{Xi}^{*}(\pi^{'*}W)\otimes \tau_{Xi}^{*}(\pi^{*}L),\;i,j=1,2
\ee
and (\ref{2.33}), (\ref{2.34}), we conclude that
\be\label{ex013}
\tau_{Xi}^{*}{\mathcal V}\cong {\mathcal V},\;i=1,2.
\ee
That is, the rank two vector bundles defined in (\ref{ex012}) are
invariant under the ${\mathbb Z}_2\times{\mathbb Z}_2$ automorphism
group of $X$ if they are constructed using the equivariant rank two
bundles $W$ and the invariant line bundles $L$ defined in
Section~\ref{Wbundles} and Section~\ref{lbundles} respectively.
If in addition $L$ is chosen to be equivariant, then we will get an
equivariant bundle ${\mathcal V}$.

Having produced invariant (respectively equivariant) rank two bundles
${\mathcal V}$ on $X$, we can proceed to construct rank four
holomorphic vector bundles $V$ on $X$ which are invariant
(respectively equivariant) under the
${\mathbb Z}_2\times{\mathbb Z}_2$ automorphism group generated by
$\tau_{Xi},\;i=1,2$. To do this, consider any two invariant rank two
vector bundles
\be\label{1.9}
V_i=\pi^{'*}W_i\otimes \pi^{*}L_i,\;i=1,2.
\ee
Note from our previous discussion that the equivariant bundles $W_i$
and, hence, $V_i$ for $i=1,2$ have non-trivial moduli spaces. For each
$V_i$, choose a specific point in its moduli space, which we also
denote by $V_i$. For this specific choice, we then construct $V$ as
extensions satisfying
\be\label{ex014}
\ses{V_1}{V}{V_2}
\ee
and consider $Ext_X^{1}(V_2,V_1)$. Then, it follows from the previous
discussion that there exists a vector subspace
\be
Ext_X^{1}(V_2,V_1)^{inv}\subseteq  Ext_X^{1}(V_2,V_1)
\ee
of complex dimension at least one, each of whose elements is ${\mathbb
Z}_2\times {\mathbb Z}_2$ invariant. That is, if $V \in
Ext_X^{1}(V_2,V_1)^{inv}$ then
\be
\tau_{Xi}^{*}V\cong V,\;i=1,2.
\ee
Therefore, any such bundle will descend to a twisted, rank four
vector bundle $V_Z$ on the non-simply connected Calabi-Yau threefold
\be
Z=X/({\mathbb Z}_2\times{\mathbb Z}_2).
\ee
Note that these conclusions hold for the choice of any points
$V_i,\;i=1,2$ in the respective moduli spaces.

The definition of $V$ as an extension of $V_2$ by $V_1$ makes the
calculation of its Chern character straightforward. It follows from
(\ref{ex014}) that
\be\label{1.11}
\ch V=\ch V_1+\ch V_2.
\ee
Using (\ref{1.9}), the Chern characters of $V_i,\; i=1,2$ can be
evaluated. They are
\be
\ch V_i=\ch\pi^{'*}W_i\cdot \ch\pi^{*}(L_i)\\
       =\pi^{'*}\ch W_i\cdot\pi^{*}\ch (L_i).\\
\ee
From  (\ref{2.30}) and  (\ref{2.31}), we find that
\be\label{1.12}
\begin{split}
\ch V_i      =&\pi^{'*}(2+(a_i+b_i)f-k_i\pt)\cdot \pi^{*}(1_{B^{'}}+c_1(L_i)+\frac{1}{2}c_1(L_i)^2)\\
       =&2+\pi^{*}(2c_1(L_i)+(a_i+b_i)f^{'})+(c_1(L_i)^2+(a_i+b_i)(c_1(L_i)\cdot f^{'}))f\times \pt^{'}\\
        & -k_i \pt\times f^{'}-k_i(c_1(L_i)\cdot f)\pt.
\end{split}
\ee
The Chern character of $V$ now follows  easily from (\ref{1.11}) and (\ref{1.12}). It is  given by
\be\label{chernV}
\begin{split}
\ch V=&4+\pi^{*}(2c_1(L_1)+2c_1(L_2)+(a_1+a_2+b_1+b_2)f^{'})\\
      &+(c_1(L_2)^2+c_1(L_2)^2+(a_1+b_1)(c_1(L_1)\cdot f^{'})+(a_2+b_2)(c_1(L_2)\cdot f^{'}))f\times \pt^{'}\\
      &-(k_1+k_2)\pt\times f^{'}
      -(k_1(c_1(L_1)\cdot f^{'})+k_2(c_1(L_2)\cdot f^{'}))\pt.
\end{split}
\ee
This formula will turn out to be  important when we try to find vector bundles that are compatible with the physical constraints discussed  in Section~\ref{physics}.

\section{Stability of Vector Bundles $V$}\label{stable}

Let $X$ be a smooth projective algebraic variety with $\dim_{\mathbb C} X =3$. A divisor  $H\in H^2(X,{\mathbb Z})$ is said to be ample if 
\be
H\cdot C >0,\;\;\;H^2\cdot S >0
\ee
for all irreducible, effective curves $C\subset X$ and all irreducible, effective surfaces $S\subset X$ and if
\be
H^3>0.
\ee
It can be shown that every projective algebraic variety admits at least one ample class. Now consider a holomorphic vector bundle or a torsion free sheaf $V$ on $X$. For the  definition of torsion free sheaves, we refer the reader to  Appendix~\ref{appendix1}.  The ``slope'' of $V$ with respect to an ample class $H$, denoted by $\mu_H(V)$, is defined to be
\be\label{slope}
\mu_H(V)=\frac{c_1(V)\cdot H^2}{\rank V}.
\ee
Then $V$ is said to be stable if there exists an ample class $H$ with respect to which 
\be\label{2.35}
\mu_H(F)<\mu_H(V)
\ee
for every torsion free sub-sheaf $F$ in $V$ with
\be\label{ex052}
\rank F < \rank V.
\ee
The notion of stability of $V$ is important for the following reason. In order for $V$ to correspond to a physical vacuum in superstring theory, it must admit a connection that satisfies the hermitian Yang-Mills equation. It was shown by Donaldson \cite{d} and Uhlenbeck and  Yau \cite{uyau} that this will be the case if and only if $V$ is holomorphic and stable. 
All of the vector bundles and torsion free sheaves $V$ discussed in this paper are holomorphic.  However, if $V$ is to be a physical vacuum, we must also demonstrate that it is stable.

In this paper, $X$ is a Calabi-Yau threefold elliptically fibered over a base $B^{'}\cong d{\mathbb P}_9$. It follows that $X$ is a projective algebraic variety and, hence, at least one ample class $H$ exists. Since $X$ is a fiber product of $B$ and $B^{'}$, it is relatively straightforward to construct a set of ample classes on $X$ from the  ample classes on $B$ and $B^{'}$. This is accomplished as follows. Let $ h_0 \in H^2(B,{\mathbb Z})$ be an ample class on $B$ and $ h_0^{'} \in H^2(B^{'},{\mathbb Z})$ be an ample class on $B^{'}$. The pull-backs, $\pi^{'*} h_0$ and $\pi^{*}h_0^{'}$, are not ample classes in $H^2(X,{\mathbb Z})$. However, it is not hard to show that their sum
\be\label{ex017}
H_0= \pi^{'*} h_0+\pi^{*}h_0^{'}  \in  H^2(X,{\mathbb Z})
\ee
is, in fact, ample in $X$. Now  consider a third ample class $h^{'} \in H^2(B^{'},{\mathbb Z})$. As stated above, $\pi^{*}h^{'}\in  H^2(X,{\mathbb Z}) $ is not ample. However,
\be\label{ex018}
H^{'}=\pi^{*}h^{'} \in  H^2(X,{\mathbb Z})
\ee
is an effective class. This is sufficient for the sum
\be\label{2.38}
H=H_0+nH^{'}\in  H^2(X,{\mathbb Z})
\ee
to be an ample class in $X$ for any integer $n \in {\mathbb Z}_{\geq 0}$. Of course, this construction supposes that we can find ample classes on $B$ and $B^{'}$. However, since $B$ and $B^{'}$ are projective algebraic varieties, such ample classes will indeed exist. Furthermore, there is an easy tool for  checking if a divisor is ample on these surfaces. This is called the Nakai-Moishezon criterion \cite{h} which states the following. A divisor $h$ on a surface is ample if and only if
\be
h^2>0,\;\;\;h\cdot C>0
\ee
for every irreducible, effective curve $C$ on the surface. This gives an easy criterion for finding ample classes on $B$ and $B^{'}$.

There may be many ample classes in $ H^2(X,{\mathbb Z})$. Why, then, do we consider those of the form (\ref{2.38})? The reason is the following. It was shown in \cite{fmw2} that vector bundles which are constructed  using smooth, irreducible spectral covers on $X$ are always stable with respect to some ample class of the form (\ref{2.38}) for sufficiently large $n$. Intuitively, $H_0$ ensures the ampleness of $H$ and $nH^{'}$ increases the volume of the fiber to ensure stability. The vector bundles $V$ in this paper, however, are not constructed from spectral data. They can indeed be so constructed, but the associated spectral cover will be reducible and non-reduced. Be that as it may, checking stability  with respect to some ample class $H$ of the form (\ref{2.38}) seems reasonable. This will be done in the next two sections where, in fact, we demonstrate the stability of $V$ with respect to ample classes of the form (\ref{2.38}).

Proving the stability of $V$ with respect to $H$ requires checking that condition (\ref{2.35}) holds for all sub-sheaves $F$ in $V$. This is technical and we will  postpone a more complete calculation until Section~\ref{Stable}. Here, however, we  present a simple criterion that is necessary in order for $V$ to be stable. This criterion will be checked, along with the conditions for obtaining standard-like model vacua, in the next section. First, note that, in this paper, we are taking the structure group of $V$ to be  $G=SU(4)$. It follows that $c_1(V)=0$ and, hence,
\be\label{2.37}
\mu_H(V)=0.
\ee
Therefore, $V$ will be stable with respect to $H$ if and only if
\be
\mu_H(F)<0
\ee
for every sub-sheaf $F$ of $V$ satisfying (\ref{ex052}). Second, although there may be many sub-sheaves of $V$, one of them is obvious and natural. Recall that $V$ is constructed as the extension
\be
\ses{V_1}{V}{V_2}.
\ee
It follows that $V_1$ is always a rank two sub-bundle of $V$. Therefore, the stability of $V$ requires, among other things, that
\be\label{2.36}
\mu_H (V_1) <0.
\ee
This constraint is by no means sufficient for the stability of $V$. However, it is very necessary and, as it turns out, very restrictive. It is helpful to check this condition first, as we will do in the next section, before demonstrating the complete stability of $V$. We will call  (\ref{2.36}) the stability condition on $V$. This criterion immediately explains why we  must consider vector bundles  $V$ that are  non-trivial extensions of $V_2$ by $V_1$. Let us assume that $V$ is the trivial extension, that is
\be\label{2.39}
V\cong V_1 \oplus V_2.
\ee
This would imply that  both $V_1$ and $ V_2$ are sub-bundles of $V$. Since the slope of $V_1$ is required to be negative by (\ref{2.36})  and the slope of $V$ is zero by (\ref{2.37}), it follows that $\mu_H(V_2)>0$, which violates (\ref{2.35}). Hence, a vector bundle $V$ of the form (\ref{2.39}) can never be stable. However, if $V$ is a non-trivial extension, $V_2$  is not a sub-bundle, and this difficulty is removed. Happily, as shown in the previous section, both $Ext^1_X(V_2,V_1)$ and $Ext^1_X(V_2,V_1)^{inv}$ have dimensions of at least one, and, hence, each contains non-trivial elements.
 If $Ext^1(V_2,V_1)$ is trivial, the best we can do is  to find $V_1$ and $ V_2$ both with vanishing slope. Then $V_1 \oplus V_2$ is potentially poly-stable. Poly-stable bundles, however, correspond to solutions of the hermitian Yang-Mills connection which are reducible. We will not consider this possibility in this paper.

\section{Numerical Conditions and Standard Model Bundles}\label{numerics}

In this section, we impose the physical conditions ${\bf(C1)}$,
${\bf(C2)}$, ${\bf(C3)}$ and the stability condition (\ref{2.36}) on
the ${\mathbb Z}_2\times {\mathbb Z}_2$ invariant, holomorphic, rank
four vector bundles constructed in Section~\ref{Vbundles}. In fact,
since none of these four conditions depends on the ${\mathbb
Z}_2\times {\mathbb Z}_2$ invariance of $V$, the results of this
section are valid for any element of $Ext^1_X(V_2,V_1)$. They apply to
the ${\mathbb Z}_2\times {\mathbb Z}_2$ invariant bundles in
$Ext^1_X(V_2,V_1)^{inv}$ by restriction.

\subsection{Numerical Conditions}

First, recall that the Chern character of a rank four vector bundle on a threefold $X$ can be expanded in terms of the Chern classes \cite{f} of $V$ as
\be\label{4.20}
\ch V=4 +c_1(V)+\frac{1}{2}(c_1(V)^2-2c_2(V))+\frac{1}{6}(c_1(V)^3-3c_1(V)\cdot c_2(V)+ 3c_3(V)).
\ee
Comparing this  expression with  (\ref{chernV}), we see that
\be
c_1(V)=\pi^{*}(2c_1(L_1)+2c_1(L_2)+(a_1+a_2+b_1+b_2)f^{'}).
\ee
Condition ${\bf(C1)}$ in (\ref{ex016}) then requires that
\be\label{4.21}
\begin{array}{c l}
(\num{C1})& 2c_1(L_1)+2c_1(L_2)=-(a_1+a_2+b_1+b_2)f^{'}.
\end{array}
\ee
After imposing this condition on $V$, it follows from (\ref{chernV}), (\ref{4.20}) and  (\ref{4.21}) that
\be\label{4.01}
\begin{split}
c_2(V) =& -(c_1(L_2)^2+c_1(L_2)^2+(a_1+b_1)c_1(L_1)\cdot f^{'}
+ (a_2+b_2)c_1(L_2)\cdot f^{'})f\times \pt^{'}\\
       &+(k_1+k_2)\pt\times f^{'}
\end{split}
\ee
and
\be\label{4.02}
c_3(V)=-2(k_1(c_1(L_1)\cdot f^{'})+k_2(c_1(L_2)\cdot f^{'}))\pt.
\ee
Recall from \cite{opr-i} that the second Chern class of the tangent bundle of the Calabi-Yau threefold $X$ is given by 
\be
c_2(TX)=12(f\times \pt^{'} + \pt\times f^{'}).
\ee
Using this and (\ref{4.01}), we see that condition ${\bf(C2)}$ in (\ref{ex016}) requires that the class
\be
\begin{array}{c r l}
 & & 12(f\times \pt^{'} + \pt\times f^{'})\\ & + &(c_1(L_1)^2+c_1(L_2)^2+(a_1+b_1)c_1(L_1)\cdot f^{'}+(a_2+b_2)c_1(L_2)\cdot f^{'})f\times \pt^{'}\\
& - & (k_1+k_2)\pt\times f^{'}
\end{array}
\ee
be effective. This leads to two  numerical constraints,
\be
\begin{array}{c l}
(\num{C2}a) & k_1+k_2\leq 12\\
\end{array}
\ee
and
\be\label{ex024}
\begin{array}{c l}
(\num{C2}b)& c_1(L_2)^2+c_1(L_2)^2+(a_1+b_1)c_1(L_1)\cdot f^{'}+(a_2+b_2)c_1(L_2\cdot f^{'})+12\geq 0.
\end{array}
\ee
The three family condition {\bf (C3)} in (\ref{ex016}) applied to (\ref{4.02}) implies that
\be\label{ex053}
\begin{array}{c l}
(\num{C3}) & k_1(c_1(L_1)\cdot f^{'})+k_2(c_1(L_2)\cdot f^{'})=-12.
\end{array}
\ee
To analyze the stability condition (\ref{2.36}) numerically, we have to calculate $\mu_H(V_1)$. Recall from the definition of $\mu_H$ given in (\ref{slope}) that to calculate the slope of any vector bundle, one must  first compute $H^2$. Assume that $H$ is of the form  (\ref{2.38}). Then, 
\be\label{H2}
\begin{split}
H^2=&(H_0+nH^{'})^2\\
%   =&H_0^2+2nH_0\cdot H^{'}+n^2H^{'2}\\
   =&(\pi^{'*} h_0+\pi^{*}h_0^{'})^2+2n(\pi^{'*} h_0+\pi^{*}h_0^{'})\cdot\pi^{*}h^{'}+n^2\pi^{*}h^{'2}\\
%   =&\pi^{'*} h_0^2+\pi^{*}h_0^{'2}+2\pi^{'*} h_0\cdot\pi^{*}h_0^{'}+2n(\pi^{'*} h_0+\pi^{*}h_0^{'})\cdot\pi^{*}h^{'}+n^2\pi^{*}h^{'2}\\
   =&h_0^2(pt\times f^{'})+2(\pi^{*}h_0^{'}+n \pi^{*} h^{'})\cdot\pi^{*'}h_0+\pi^{*}(h_0^{'}+nh^{'})^2,\\
\end{split}
\ee
where we have used (\ref{ex017}) and (\ref{ex018}). Second, note that  the first Chern class of the rank two bundle $V_1$ can be extracted from  (\ref{1.12}). It is given by
\be
c_1(V_1)=\pi^{*}(2c_1(L_1)+(a_1+b_1)f^{'}).
\ee
Using this, expression   (\ref{H2}), $\rank V_1=2$ and the facts that 
\be
\pi^{*}(2c_1(L_1)+(a_1+b_1)f^{'})\cdot H^2=(2c_1(L_1)+(a_1+b_1)f^{'})\cdot\pi_{*}H^2
\ee
and
\be
\pi_{*}\pi^{'*}h_0=h_0\cdot f + p \pt,\;p \in {\mathbb Z}_{\geq 0},
\ee
we find that 
\be
\mu_H(V_1)=\frac{1}{2}(2c_1(L_1)+(a_1+b_1)f^{'})\cdot(h_0^2f^{'}+2h_0^{'}(h_0\cdot f))+n(2c_1(L_1)+(a_1+b_1)f^{'})\cdot h^{'}(h_0\cdot f).
\ee
%\be
%\begin{split}
%2\mu_H(V_1)=&\pi^{*}(2c_1(L_1)+(a_1+b_1)f^{'})\cdot H^2\\
 %          =&(2c_1(L_1)+(a_1+b_1)f^{'})\cdot\pi_{*}H^2\\
 %          =&(2c_1(L_1)+(a_1+b_1)f^{'})\cdot(h_0^2f^{'}+2(h_0^{'}+nh^{'})\cdot\%pi_{*}\pi^{'*}h_0+k\pt)\\
%           =&(2c_1(L_1)+(a_1+b_1)f^{'})\cdot(h_0^2f^{'}+2h_0^{'}(h_0\cdot f))+2n(2c_1(L_1)+(a_1+b_1)f^{'})\cdot h^{'}(h_0\cdot f).\\
%\end{split}
%\ee
Recall that we want to obtain stable bundles for large values of $n$. Clearly, for large $n$ the  term $n(2c_1(L_1)+(a_1+b_1)f^{'})\cdot h^{'}(h_0\cdot f)$ will be dominant. Furthermore, $h_0\cdot f>0$ by the ampleness assumption on $h_0$. Therefore, we find that the stability condition (\ref{2.36}) requires that
\be\label{ex027}
(2c_1(L_1)+(a_1+b_1)f^{'})\cdot h^{'} <0.
\ee
Before solving these constraints, we will make one arbitrary  choice. Intersect $(\num{C1})$ in (\ref{4.21}) with $f^{'}$. Since $f^{'2}=0$, we find that
\be\label{4.1}
c_1(L_1)\cdot f^{'}+c_1(L_2)\cdot f^{'}=0.
\ee
Assuming $c_1(L_1)\cdot f^{'}=0$ would imply  that $c_1(L_2)\cdot f^{'}=0$, in contradiction to  $(\num{C3})$ in (\ref{ex053}). Therefore, we need to require that $c_1(L_1)\cdot f^{'}\neq0$. Clearly, either
\be
c_1(L_1)\cdot f^{'}>0
\ee
and, hence, from (\ref{4.1})  that 
\be\label{4.2}
c_1(L_1)\cdot f^{'}>c_1(L_2)\cdot f^{'},
\ee
or
\be
c_1(L_1)\cdot f^{'}<0
\ee
implying
\be\label{ex054}
c_1(L_1)\cdot f^{'}<c_1(L_2)\cdot f^{'}.
\ee
Henceforth, we will choose condition (\ref{4.2}). This choice is motivated by the calculation in \cite{dopw-ii} where (\ref{4.2}) is imposed to ensure the existence of non-trivial elements in $Ext^1_X(V_2,V_1)$. We will show  in \cite{dopr-ii} that this choice also ensures the existence of non-trivial elements of $Ext^1_X(V_2,V_1)$ in the context of this paper. Choosing condition (\ref{ex054}) would not allow us to arrive at this important conclusion.

In the next sub-section, we will  solve all of the constraints  $(\num{C1})$, $(\num{C2}a)$,  $(\num{C2}b)$, $(\num{C3})$, (\ref{ex027}) and (\ref{4.2}) simultaneously, thus proving the existence of rank four  vector bundles that are consistent with particle physics phenomenology.

\subsection{Solution}

In this sub-section, we will construct rank four vector bundles which
solve all of the numerical conditions described above. We begin as
follows. Note that (\ref{4.1}) and (\ref{4.2}) imply
\be\label{ex020}
0<c_1(L_1)\cdot f^{'}=-c_1(L_2)\cdot f^{'}.
\ee
Using this result in $(\num{C3})$ gives
\be\label{ex021}
(k_2-k_1)c_1(L_1)\cdot f^{'}=12.
\ee
Let us define 
\be
k\equiv k_2-k_1.
\ee
Since $c_1(L_1)\cdot f^{'}$ is positive, it follows from (\ref{ex021}) that
\be\label{ex043}
k>0
\ee
and, hence,
\be\label{ex044}
k_1<k_2.
\ee
Furthermore, since we will ensure that $L_1$ is a line bundle, the intersection $c_1(L_1)\cdot f^{'}$ is integral.  It then  follows from this and (\ref{ex021}) that
\be\label{ex045}
k \;\text{divides}\; 12.
\ee
Now, recall from (\ref{ex023}), that $k_1$ and $k_2$ must be of the form
\be\label{ex046}
k_i=4m_i+2n_i,\;m_i \in {\mathbb Z}_{\geq 0},\;n_i=0,\dots,8,\;i=1,2.
\ee
Clearly, (\ref{ex043}),(\ref{ex044}),(\ref{ex045}),(\ref{ex046}) and $(\num{C2}a)$ put severe restrictions on the allowed values of $k, k_1$ and $k_2$. In fact, there is only a small, finite set of solutions, which we list in the first three columns of Table~\ref{4.3}. 
\begin{table}
\begin{center}
\begin{tabular}{|c|c|c||c|c|c|} \hline
$k$ & $k_{1}$ & $k_{2}$ & $k_1+k_2$ & $c_1(L_{1})\cdot f'$ & $c_1(L_{2})\cdot f'$ \\ \hline\hline
$2$ & $0$ & $2$ & $2$ & $6$ & $-6$  \\ \hline
$4$ & $0$ & $4$ & $4$ & $3$ & $-3$  \\ \hline
$6$ & $0$ & $6$ & $6$ & $2$ & $-2$  \\ \hline
$2$ & $2$ & $4$ & $6$ & $6$ & $-6$  \\ \hline
$4$ & $2$ & $6$ & $8$ & $3$ & $-3$  \\ \hline
$6$ & $2$ & $8$ & $10$ & $2$ & $-2$  \\ \hline
$2$ & $4$ & $6$ & $10$ & $6$ & $-6$  \\ \hline
$4$ & $4$ & $8$ & $12$ & $3$ & $-3$  \\ \hline
$12$ & $0$ & $12$ & $12$ & $1$ & $-1$  \\ \hline
\end{tabular}
\end{center} 
\caption{Allowed values for $k,k_{1}$, $k_{2}$, $c_1(L_{1})\cdot f'$,
$c_1(L_{2})\cdot f'$}
\label{4.3}
\end{table}
The fourth column ensures that condition $(\num{C2}a)$ is obeyed. The
fifth and sixth columns give the corresponding intersection numbers of
$c_1(L_1)\cdot f^{'}$ and $c_1(L_2)\cdot f^{'}$ which solve
(\ref{ex020}) and (\ref{ex021}).

To proceed, recall that $L_i,\;i=1,2$ must be invariant line bundles
on $B^{'}$. It was shown in Section~\ref{lbundles} that, for this to
be the case, their first Chern classes must satisfy
\be\label{4.03}
c_1(L_i)=\tilde{a}_ii^{'}+x_if^{'}+y_i(n_1^{'}+o^{'}_2)+\bar{a}_iM^{'},\;i=1,2.
\ee
In the following calculation, we will make frequent use of the intersection numbers of the classes $i^{'},f^{'},n_1^{'}+o^{'}_2 $ and $M^{'}$ which were given in Table~\ref{1.6}. The convenience of this choice of generators, with these specific intersection numbers, will now become clear. Intersecting  $c_1(L_i),\;i=1,2$ with $f^{'}$, and using (\ref{4.03}), we find that
\be
c_1(L_1)\cdot f^{'}=4\tilde{a}_1,\;\;\;\;c_1(L_2)\cdot f^{'}=4\tilde{a}_2.
\ee
Note that conditions (\ref{ex020}) and (\ref{ex021})  then fix the form of $\tilde{a}_1$ and $\tilde{a}_2$. We obtain
\be\label{ex031}
\tilde{a}_1=\frac{3}{k},\;\;\;\;\tilde{a}_2=-\frac{3}{k}.
\ee
It follows that the line bundles $L_i,\;i=1,2$ are restricted to satisfy
\be\label{4.4}
\begin{array}{c l}
c_1(L_1)&=\frac{3}{k}i^{'}+x_1f^{'}+y_1(n_1^{'}+o_2^{'})+\bar{a}_1M^{'}\\
\end{array}
\ee
and
\be\label{ex022}
\begin{array}{c l}
c_1(L_2)&=-\frac{3}{k}i^{'}+x_2f^{'}+y_2(n_1^{'}+o_2^{'})+\bar{a}_2M^{'}.
\end{array}
\ee
Note that choosing $k_1, k_2 $ and $k$ from Table~\ref{4.3} and line bundles satisfying  (\ref{4.4}) and (\ref{ex022}) solves the constraints  $(\num{C2}a)$,  $(\num{C3})$ and (\ref{4.2})  completely. It also solves part of the condition $(\num{C1})$. To complete  $(\num{C1})$, we have to impose the constraints
\be\label{4.5}
\begin{split}
             2x_1+2x_2=&-(a_1+a_2+b_1+b_2),\\
             y_1=&-y_2,\\
        \bar{a}_1=&-\bar{a}_2.
\end{split}
\ee
If the coefficients $x_i, y_i $ and $\bar{a}_i$ of $c_1(L_i),\;i=1,2$ fulfill the conditions in (\ref{4.5}), then the vector bundle $V$ has vanishing first Chern class and, hence,  solves $(\num{C1})$.

To solve constraint $(\num{C2}b)$  in (\ref{ex024}), we need to compute $c_1(L_i)^2,\;i=1,2$. Using the expressions (\ref{4.4}), (\ref{ex022}), (\ref{4.5}) and Table~\ref{1.6}, we find that
%\be\label{4.05}
%\begin{split}
%c_1(L_1)^2=&\frac{9}{k^2}-4-4y_1^2+\bar{a}_1^2M^2+2(4\frac{3x_1}{k}+4\frac{3}{k}y_1)\\
%     =&-\frac{36}{k^2}-4y_1^2+\bar{a}_1^2M^2+\frac{24}{k}(x_1+y_1)\\
%c_1(L_2)^2=&\frac{9}{k^2}-4-4y_2^2+\bar{a}_2^2M^2+2(-4\frac{3x_2}{k}-4\frac{3}{k}y_2)\\
%     =&-\frac{36}{k^2}-4y_2^2+\bar{a}_2^2M^2-\frac{24}{k}(x_2+y_2).\\
%\end{split}
%\ee
\be\label{4.05}
c_1(L_1)^2=-\frac{36}{k^2}-4y_1^2-4\bar{a}_1^2+\frac{24}{k}(x_1+y_1)
\ee
and
\be\label{ex025}
c_1(L_2)^2=-\frac{36}{k^2}-4y_2^2-4\bar{a}_2^2-\frac{24}{k}(x_2+y_2).
\ee
Constraint $(\num{C2}b)$ can now be written as
\be
\begin{split}
-\frac{72}{k^2}-4y_1^2-4y_2^2-4(\bar{a}_1^2+\bar{a}_2^2)+\frac{24}{k}(x_1+y_1-x_2-y_2)+\frac{12}{k}(a_1+b_1-a_2-b_2)\geq -12.\\
\end{split}
\ee
After implementing the conditions (\ref{4.5}) and completing the squares,  we find that
\be\label{e1}
\begin{split}
-8(y-\frac{3}{k})^2-8\bar{a}^2+\frac{24}{k}(x_1-x_2)+\frac{24}{k}(a_1+b_1+x_1+x_2)\geq -12,\\
\end{split}
\ee
where we have defined
\be\label{ex028}
y \equiv y_1= -y_2,\;\;\;\bar{a}\equiv \bar{a}_1=-\bar{a}_2.
\ee
Furthermore, defining 
\be\label{ex029}
u\equiv 2y-\frac{6}{k},\;\;\;\;\;x\equiv 2x_1+a_1+b_1
\ee
allows us to simplify condition $(\num{C2}b)$ in (\ref{e1}) to
\be\label{4.6}
-u^2-4\bar{a}^2+\frac{12}{k}x \geq -6.
\ee
The expressions (\ref{4.4}) and (\ref{ex022}) for  $c_1(L_i),\;i=1,2$ written in terms of $x,u$ and $\bar{a}$ are
\be\label{4.7}
\begin{split}
c_1(L_1)=&\frac{3}{k}i^{'}+\frac{1}{2}(x-a_1-b_1)f^{'}+\frac{1}{2}(u+\frac{6}{k})(n_1+o_2)+\bar{a}M^{'}\\
\end{split}
\ee
and
\be\label{ex026}
\begin{split}
c_1(L_2)=&-\frac{3}{k}i^{'}+\frac{1}{2}(-x-a_2-b_2)f^{'}+\frac{1}{2}(-u-\frac{6}{k})(n_1+o_2)-\bar{a}M^{'}.
\end{split}
\ee
Note that if $u,\bar{a}$ and $k$ obey the relations imposed by condition (\ref{4.6}), then the vector bundle $V$, defined in terms of line bundles $L_i,\;i=1,2$ satisfying (\ref{4.7}), solves both  constraints $(\num{C1})$ and $(\num{C2}b)$. Finally, note that if we use the expansion for $c_1(L_1)$ given in (\ref{4.7}), then the stability condition  (\ref{ex027}) can be written as 
\be\label{4.8}
(\frac{6}{k}i^{'}+xf^{'}+(u+\frac{6}{k})(n_1^{'}+o_2^{'})+2\bar{a}M)\cdot h^{'}<0.
\ee

To complete our numerical solution, we have to find values of $u,\bar{a},k,x,a_1,b_1,a_2,b_2$ for which equations (\ref{4.6}) and (\ref{4.8}) are satisfied, and such that the expressions  $c_1(L_i),\;i=1,2$ in (\ref{4.7}) are integral. That is, such that  $L_i,\;i=1,2$ are indeed line bundles.
The solution we are presenting in this paper is specified by
\be\label{4.11}
u=-1,\;\;\;\bar{a}=\frac{1}{2},\;\;\;k=6,\;\;\;x=-1,\;\;\;a_i+b_i \in 2{\mathbb Z},\;i=1,2.
\ee
Other, less obvious, solutions may exists. But we  will not pursue them here.
Using (\ref{ex031}), (\ref{4.5}), (\ref{ex028}) and (\ref{ex029}), we see that this solution corresponds to taking 
\be
\tilde{a}_1=-\tilde{a}_2=\frac{1}{2},\;\;\;y_1=y_2=0,\;\;\;\bar{a}_1=-\bar{a}_2=\frac{1}{2}
\ee
and
\be
x_1=\frac{-a_1-b_1-1}{2},\;\;\;x_2=\frac{-a_2-b_2+1}{2}
\ee
with $a_i+b_i \in 2{\mathbb Z},\;i=1,2$. In addition, we see from Table~\ref{1.6} that $k=6$ can occur in two ways, either 
\be\label{ex030}
k_1=2,\;\;\;k_2=8
\ee
or
\be\label{ex032}
k_1=0,\;\;\; k_2=6.
\ee
Note, however, that (\ref{ex032}) implies that $V_1$ is isomorphic to the  sum of two line bundles.  Therefore, vector bundles $V$ corresponding to (\ref{ex032}) are not irreducible. Hence,  we discard them and allow  (\ref{ex030}) only.
For the solution presented in (\ref{4.11}), the left hand side of  condition (\ref{4.6}) reads
\be
-u^2-4\bar{a}^2+\frac{12}{k}x=-4 \geq -6,
\ee
which satisfies this constraint. To proceed, it is convenient to  define a divisor $D^{'}$ given by 
\be\label{ex033}
D^{'}\equiv \frac{1}{2}(i^{'}-f^{'}+2\bar{a}M^{'})=e_2^{'}+e_4^{'}+e_6^{'}-e_9^{'}-n_1^{'}-n_2^{'}.
\ee
 Note that the first Chern class of the line bundles $L_i,\;i=1,2$ can  be written in terms of this divisor.  We find that
\be\label{4.9}
c_1(L_1)=D^{'}-\frac{1}{2}(a_1+b_1)f^{'},\;\;\;c_1(L_2)=-D^{'}-\frac{1}{2}(a_2+b_2)f^{'}.
\ee
Since the divisor $D^{'}$ is clearly integral, the Chern classes
defined in (\ref{4.9}) are also integral for $a_i+b_i \in 2{\mathbb
Z},\;i=1,2$. Hence $L_i,\;i=1,2$ are line bundles, as required. By
construction the line bundle $\oz(D')$  and, hence,  the line bundles
$L_{1}$ and $L_{2}$ are   ${\mathbb Z}_{2}\times {\mathbb Z}_{2}$
invariant. However $\oz(D')$ can not be equivariant. Indeed, if it is,
then its restriction to the fiber $f'_{0}$, that is, the fiber $f^{'}$ over the point $0\in \cp{1}$,  will also be
equivariant. On the other hand, the group ${\mathbb Z}_{2}\times
{\mathbb Z}_{2}$ preserves $f'_{0}$ and acts on it as translations by
points of order two. But $D'\cdot f' = 2$. Therefore,  $\oz(D')|_{f'_{0}}$
can not be a pull-back from $f'_{0}/({\mathbb Z}_{2}\times {\mathbb
Z}_{2})$ since, in that case,  $D'\cdot f'$ would  have to be divisible by
four.

Using (\ref{ex033}), constraint (\ref{4.8}) becomes the condition that 
\be\label{4.12}
D^{'}\cdot h^{'} < 0
\ee
for some ample class $h^{'}$ on $B^{'}$.
To solve this condition, we must  find such an ample class. 
The existence of this  class is by no means obvious. For example, if $D^{'}$ were to  be an effective divisor and, hence, consist of irreducible, effective curves in  $B^{'}$, we would have $D^{'}\cdot h^{'} >0  $ for all ample classes $h^{'}$. This follows from the Nakai-Moishezon criterion discussed previously. Therefore, it is necessary to show that $D^{'}$ in (\ref{ex033}) is not effective. In other words, we need to show, that $H^{0}(B^{'},\oz(D^{'}))=0$. Happily, this is  easy to do. Let us assume that $D^{'}$ is an effective divisor in $B^{'}$. Furthermore, assume that the intersection of $D^{'}$  with some other effective divisor $e^{'}$ is negative. Then the divisor $D^{'}-e^{'}$ is effective. That this is true can be seen as follows.  Let us write the short exact sequence
\be\label{4.10}
\ses{\oz(D^{'}-e^{'})}{\oz(D^{'})}{{\mathcal O}_{e^{'}}(-n)}.
\ee
Here $\oz(D^{'}-e^{'})$ denotes the sheaf of all sections of the line bundle $\oz(D^{'})$ which vanish along the curve $e^{'}$ and  ${\mathcal O}_{e^{'}}(-n)$ is a line bundle on $e^{'}$ of negative degree where  $-n=D^{'}\cdot e^{'},\; n \in {\mathbb Z}_{>0}$. Each short exact sequence of sheaves implies a long exact sequence in sheaf cohomology.  The long exact sequence associated with (\ref{4.10}) is given by
\be\label{4.04}
0 \to H^0(B^{'},\oz(D^{'}-e^{'})) \to H^0(B^{'},\oz(D^{'})) \to H^0(B^{'},{\mathcal O}_{e^{'}}(-n))\to \dots
\ee
Since ${\mathcal O}_{e^{'}}(-n)$ has negative degree, $H^0(B^{'},{\mathcal O}_{e^{'}}(-n))=0$. Hence, (\ref{4.04}) implies that
\be
H^0(B^{'},\oz(D^{'}-e^{'})) \cong H^0(B^{'},\oz(D^{'})).
\ee
Since $H^0(B^{'},\oz(D^{'}))\neq 0$ by assumption, it follows that  $H^0(B^{'},\oz(D^{'}-e^{'}))\neq 0$. Hence, the divisor $D^{'}-e^{'}$ is effective, as claimed above. Now take $e^{'}$ to be any one of the divisors   $e_i,i=2,4,6$. Recall that these are a subset of the nine divisors on $B^{'}$ with self-intersection $-1$ discussed in Section~\ref{Z}.
Assume that $D^{'}$ is effective. Since $D^{'} \cdot e_2^{'}=-1$, where we have used the  intersection numbers  given in \cite{opr-ii}, the divisor $D^{'}-e_2^{'}$ has to be effective. Furthermore, since $(D^{'}-e_2^{'})\cdot e_4^{'}=-1$, the divisor $D^{'}-e_2^{'}-e_4^{'}$ must also be effective. Finally, using $(D^{'}-e_2^{'}-e_4^{'})\cdot e_6^{'}=-1$, we find that the divisor
\be
D^{'}-e_2^{'}-e_4^{'}-e_6^{'}=-e_9^{'}-n_1^{'}-n_2^{'}
\ee
is also effective. But this is  clearly impossible, since $e_9^{'},n_1^{'}$ and $n_2^{'}$ are independent effective divisors. Hence, we have shown that $D^{'}$ can not be effective, as required. 
The same reasoning can be applied to show that any integral multiple  $nD^{'}$, for positive integer $n$, can not be effective. One can now Kleinmann's criterion for ampleness to conclude that there must exist an  ample class $h^{'}$ such that 
\be\label{a1}
D^{'}\cdot h^{'}<0.
\ee
Therefore, we have shown that condition (\ref{4.12}) can always be satisfied. If follows that we have found vector bundles $V$ on $X$ which fulfill all of the numerical conditions.

Before concluding this section, let us make two final observations. First, the above argument clearly breaks down for the divisor 
\be
D^{'}+f^{'}.
\ee 
This follows simply from the fact that $f^{'}\cdot e_i=1 $ for $i=1,\dots,9$ and, hence, that  $(D^{'}+f^{'})\cdot e_i=0$ for $i=2,4,6$. It can indeed be shown \cite{dopr-ii} that $D^{'}+f^{'}$ is effective. This fact puts a  severe restriction on the possible solutions of the numerical conditions, since it makes (\ref{4.8}), in conjunction with (\ref{4.6}) and the integral conditions  (\ref{4.7}) and (\ref{ex026}), hard to solve. Be that as it may, there may be, in addition to (\ref{4.11}), other  solutions of the numerical constraints. We will not pursue them in this paper. 

The second observation is of fundamental importance in our construction. Note that the argument leading to expression (\ref{a1}) required the existence of two effective divisors, $n_1^{'}$ and $n_2^{'}$, in $D^{'}$. What is the origin of these two classes? Recall that for the restricted two parameter family of rational elliptic surfaces $B^{'}$ that we are considering in this paper, $H_2(B^{'},{\mathbb Z})^{inv}\otimes {\mathbb Q}$ is generated by the invariant classes $i^{'},f^{'},n_1^{'}+o_2^{'}$ and $M^{'}$, where $i^{'}$ and $M^{'}$ are defined in (\ref{b.6}). Note that $n_1^{'}$ enters $i^{'}$ and $M^{'}$, as well as $n_1^{'}+o_2^{'}$, whereas $n_2^{'}$ appears in $M^{'}$. Both $n_1^{'}$ and $n_2^{'}$  are irreducible components of some $I_2$ fibers and, hence, are effective. These two classes enter $D^{'}$ through its definition (\ref{ex033}) as an element of $H_2(B^{'},{\mathbb Z})^{inv}\otimes {\mathbb Q}$. What would happen if instead of the two parameter family, we considered the less restrictive three parameter family  of rational elliptic surfaces $B^{'}$? In this case, it is not too hard to show that the  rank of $H_2(B^{'},{\mathbb Z})^{inv}\otimes {\mathbb Q}$ is the same, that is, four. In addition, the structure of its generators is similar to the above but with the important difference that $n_1^{'}$ and $n_2^{'}$ are now replaced  by two different classes, which we denote by $l_1^{'}$ and $l_2^{'}$. These new classes are not components of $I_2$ fibers and, crucially, are not effective. It follows that the argument leading to expression (\ref{a1}) will no longer be valid. Hence, for generic surfaces $B^{'}$ in the three parameter family, we are unable to satisfy the necessary constraints 
${\bf(C1)}$, ${\bf(C2)}$, ${\bf(C3)}$ and (\ref{2.36}). This explains, at long last, why we have required $B^{'}$ to be in the restricted two parameter family of rational elliptic surfaces.

To conclude  this section, let us state what has been  achieved so far.  We have constructed  rank four holomorphic vector bundles  $V$ as extensions
\be\label{5.0}
\ses{V_1}{V}{V_2}
\ee
of  $V_2$ by $V_1$, where
\be\label{ex034}
V_1=\pi^{'*}W_1\otimes \pi^{*}L_1,\;\;\;V_2=\pi^{'*}W_2\otimes \pi^{*}L_2
\ee
are ${\mathbb Z}_2\times {\mathbb Z}_2$ invariant. The rank two vector bundles $W_i,\;i=1,2$ are taken to be locally free extensions of rank one torsion free sheaves on $B$,
\be\label{5.1}
\ses{\oy(a_if)}{W_i}{\oy(b_if)\otimes I_{z_{k_i}}},\;i=1,2
\ee
with $k_1=2$, $k_2=8$ and $a_i+b_i \in 2{\mathbb Z}$ for both $i=1,2$.  The  line bundles $L_i,\;i=1,2$ have first Chern class  of the form
\be
c_1(L_1)=D^{'}-\frac{1}{2}(a_1+b_1)f^{'},\;\;\;c_1(L_2)=-D^{'}-\frac{1}{2}(a_2+b_2)f^{'}
\ee
with the divisor $D^{'}$  given by
\be
D^{'}=e_2^{'}+e_4^{'}+e_6^{'}-e_9^{'}-n_1^{'}-n_2^{'}.
\ee
Any rank four vector bundle $V$ in $Ext^1(V_2,V_1)$ fulfills all of the numerical conditions $(\num{C1})$, $(\num{C2}a)$,  $(\num{C2}b)$, $(\num{C3})$, (\ref{ex027}) and (\ref{4.2}). However, as stated above, demonstrating that condition (\ref{ex027}) is satisfied is a necessary, but not sufficient,  condition for $V$ to be stable. Therefore, in the next section, we analyze the remaining conditions required to prove the stability of $V$.

\section{Proof of the Stability of $V$}\label{Stable}

In the previous section, we presented rank four  holomorphic vector bundles $V$ that satisfy all the constraints ${\bf(C1)}$, ${\bf(C2)}$ and   ${\bf(C3)}$ required by particle physics phenomenology. In addition, we showed that these bundles also satisfy an important necessary condition for stability, namely (\ref{2.36}).
In this section, we will prove the stability of the rank four holomorphic vector bundles  constructed in the previous section for a large class of torsion free sub-sheaves. More precisely, we will show that all torsion free sub-sheaves of  $V$, whose restriction to the generic fiber $f$ of  $\pi: X \to B^{'}$ have negative degree, can never destabilize $V$. The remaining relevant torsion free sub-sheaves, those whose restriction to a generic fiber have vanishing degree, are much harder to analyze. This analysis will be carried out in \cite{dopr-ii} where it will be shown that, subject to a mild further restriction on $V$, these sub-sheaves also can never destabilize $V$. We will simply state this final restriction at the end of this section.  As in Section~\ref{numerics}, nowhere in this section will we require that $V$ be ${\mathbb Z}_2\times {\mathbb Z}_2$ invariant, only that $V_1$ and $V_2$ are. It follows that our results will apply to any element in $Ext^{1}_X(V_2,V_1)$ and, hence, to the invariant bundles in $Ext^{1}_X(V_2,V_1)^{inv}$ by restriction.

We  begin with the  observation that if  $F$ is a torsion free sub-sheaf of $V$, which we denote by
\be
F \subset V,
\ee
then the degree of $F$ restricted to a fiber $f$ of the elliptic fibration $\pi: X \to B^{'}$ is non-positive. Note that we have shown in \cite{opr-i} using the fiber product structure $X=B\times_{\cp{1}} B^{'}$ that the fiber of $\pi: X \to B^{'}$ can be identified with the  fiber of $\beta: B \to \cp{1}$. It is consistent, therefore, to denote both fibers by $f$. Then, we claim that
\be\label{s.1}
c_1(F|_f)\leq 0.
\ee
Also, note  that 
\be
F|_f \subset V|_f.
\ee
The proof of (\ref{s.1}) proceeds in several steps. We begin by  restricting  the sequence (\ref{5.0}) to a smooth fiber $f$. Then,  $V|_f$ fits into the short exact sequence
\be\label{ex036}
\ses{V_1|_f}{V|_f}{V_2|_f}.
\ee
Let us analyze $V_1|_f$. First, using (\ref{ex034}), note that 
\be
V_1|_f\cong (\pi^{'*}W_1 \otimes \pi^{*}L_1)|_f\cong \pi^{'*}W_1|_f \otimes \pi^{*}L_1|_f.
\ee
Since $\pi^{*}L_1$ is the pull-back of a line bundle under $\pi: X \to B^{'}$ and we choose $f $ to be a generic fiber of $\pi$, then the restriction $\pi^{*}L_1|_f$ has degree zero. Furthermore, it is isomorphic to the trivial bundle ${\mathcal O}_f$ on the fiber $f$. That is,
\be
\pi^{*}L_1|_f \cong {\mathcal O}_f.
\ee
Hence, 
\be\label{s.2}
V_1|_f\cong\pi^{'*}W_1|_f.
\ee
Note from the identification of the fibers of $X$ and $B$ that
\be\label{s.3}
\pi^{'*}W_1|_f \cong W_1|_f.
\ee
If we restrict the defining sequence (\ref{5.1}) of $W_1$ to a generic smooth fiber $f$ which does not contain points in the ideal sheaf $I_{z_{2}}$, we obtain
\be
\ses{{\mathcal O}_f}{W_1|_f}{{\mathcal O}_f}.
\ee
Hence, $W_1|_f$ corresponds to an element of $Ext^{1}_f({\mathcal O}_f,{\mathcal O}_f)$. We can calculate this group using sheaf cohomology. Note, using  (\ref{1.20}), that
\be
Ext^{1}_f({\mathcal O}_f,{\mathcal O}_f)\cong H^1(f,{\mathcal O}_f).
\ee
Then, Serre duality on the smooth genus one curve $f$ implies
\be
H^1(f,{\mathcal O}_f)\cong H^0(f,K_f),
\ee 
where $K_f$ denotes the canonical  bundle on the torus $f$. Since this bundle is trivial, it has one global section. Therefore,
\be
Ext^{1}_f({\mathcal O}_f,{\mathcal O}_f)\cong {\mathbb C}
\ee
and $W_1|_f$ could, in principle,  correspond to a non-trivial extension. However, as we will show in \cite{dopr-ii}, $W_1|_f$ must correspond to the trivial extension and, hence,
\be\label{ex035}
W_1|_f\cong  {\mathcal O}_f \oplus {\mathcal O}_f.
\ee
It follows from (\ref{s.2}), (\ref{s.3}) and (\ref{ex035})that
\be
V_1|_f\cong  {\mathcal O}_f \oplus {\mathcal O}_f.
\ee
In a similar manner, one can  show that $V_2|_f\cong  {\mathcal O}_f \oplus {\mathcal O}_f$. Therefore, the restricted sequence (\ref{ex036}) becomes
\be\label{ex037}
\ses{{\mathcal O}_f \oplus {\mathcal O}_f}{V|_f}{{\mathcal O}_f \oplus {\mathcal O}_f}.
\ee

Using this sequence, we now continue the proof of (\ref{s.1}). As the next step, consider rank one sub-bundles $F \subset V$.  Recall that $F|_f \subset V_f$. From the diagram
\be
{\xymatrix{
0 \ar[r] & {{\mathcal O}_f \oplus {\mathcal O}_f} \ar[r]^-{u} & {V|_f} \ar[r]^-{v} & {{\mathcal O}_f \oplus {\mathcal O}_f} \ar[r] & 0 \\  
         &            &  {F|_f}\ar[u]^-{i} \ar@{.>}[ul]^{\gamma} \ar@{.>}[ur]_{\lambda}  &            & \\
         &            &  0  \ar[u] &            &
}}
\ee
where $i$ is  inclusion, we see that there is a composition map $\lambda=v \circ i$ such that
\be\label{5.2}
\lambda: F|_f \to  {\mathcal O}_f \oplus {\mathcal O}_f.
\ee
Let us assume  this map is zero, that is, $\lambda=0$. This assumption  implies that the image $i(F|_f)\subset V|_f$ maps to zero under the quotient map $v:  V|_f \to {\mathcal O}_f \oplus {\mathcal O}_f$. Since (\ref{ex037}) is exact, the image $i(F|_f)\subset V|_f$ has a pre-image $u^{-1}(i(F|_f))$ in ${\mathcal O}_f \oplus {\mathcal O}_f$. Hence,  there exists a map $\gamma=u^{-1}\circ i$ such that
\be
\gamma: F|_f  \to {\mathcal O}_f \oplus {\mathcal O}_f.
\ee
Projecting the image of $\gamma$ into either of these two factors gives us a map $\bar{\gamma}:  F|_f \to {\mathcal O}_f$.  Then, either  $F|_f$ has negative degree or $F|_f \cong {\mathcal O}_f$.
That this is true can easily be seen. Any holomorphic map 
\be
\bar{\gamma}:  F|_f \to {\mathcal O}_f
\ee
is an element of $Hom_f(F|_f,{\mathcal O}_f)$. Equivalently, since
\be
Hom_f(F|_f,{\mathcal O}_f)\cong H^0(f,F^{*}|_f \otimes{\mathcal O}_f ),
\ee
$\bar{\gamma}$ corresponds to a global section of $F^{*}|_f$. But if $F|_f$ has positive degree, then $F^{*}|_f$ has negative degree. Hence, there are no global sections of $F^{*}|_f$ and $\bar{\gamma}$ does not exist, in contradiction to the above result. Hence $\deg(F|_f)\leq 0$. Let us consider the case when this degree  vanishes. But a line bundle of degree zero which has a global section is the trivial line bundle, hence $F|_f \cong {\mathcal O}_f$.

What happens if the map $\lambda$ in (\ref{5.2}) is not the zero map? We now show that, in this case, $\lambda$ must to be injective, that is,
\be\label{inc}
\ker \lambda=0.
\ee
To prove this,  assume $\lambda$ is not injective. Then, $\ker \lambda$ is a non-trivial sub-sheaf of $F|_f$ and, hence, is of rank zero or rank one. If $\ker \lambda$ has rank zero, then it would be torsion in $F_f$,
\be
0\to \ker\lambda \to F|_f.
\ee
But this is impossible, since we assumed $F|_f$ to be locally free and, hence, torsion free. If $\ker\lambda$ is of rank one, then the image $\lambda(F|_f)$ in ${\mathcal O}_f\oplus {\mathcal O}_f$ must be of rank zero. Therefore,
\be
0 \to  \lambda (F|_f) \to {\mathcal O}_f\oplus {\mathcal O}_f
\ee
would imply that $\lambda (F|_f)$ is a torsion sub-sheaf of ${\mathcal O}_f\oplus {\mathcal O}_f$, which is again impossible. Hence, we have shown that if $\lambda$ is not the zero map then $\ker \lambda$ vanishes. It follows that 
\be
\lambda: F|_f \to {\mathcal O}_f\oplus {\mathcal O}_f
\ee
is an injection and that $F|_f$ a sub-sheaf of ${\mathcal O}_f\oplus {\mathcal O}_f$. We can now argue as above, proving that $F|_f$ has non-positive degree. The same argument works for  all torsion free sub-sheaves of  any rank. Therefore, we can make the important assertion that if  $F$ is any torsion free sub-sheaf of $V$, then the degree of $F|_f$ is  non-positive and (\ref{s.1}) is proven.

We will now prove the stability of $V$ with respect to all torsion free sub-sheaves $F$ with strictly negative degree, that is, 
\be\label{5.3}
c_1(F|_f)<0.
\ee
More precisely, we will  show that for any  vector bundle $V$ defined from sequence  (\ref{5.0}), there is an ample class $H$ with respect   to which  each torsion free  sub-sheaf satisfying (\ref{5.3}) has negative slope. Since the slope of $V$ vanishes, it follows that these sub-sheaves can never destabilize $V$. As discussed previously, we choose $H$ of the form
\be\label{s.6}
H=H_0+nH^{'} \in  H^2(X,{\mathbb Z})
\ee 
where $n$ is a non-negative integer,
\be
H_0= \pi^{'*} h_0+\pi^{*}h_0^{'} 
\ee
and 
\be
H^{'}=\pi^{*}h^{'}.
\ee
Here $ h_0 \in H^2(B,{\mathbb Z})$ denotes an ample class in $B$ and $ h_0^{'}, h^{'} \in H^2(B^{'},{\mathbb Z})$ are two  ample classes in $B^{'}$. The motivation for choosing the ample class $H$ to be of the form (\ref{s.6}) was given in Section~\ref{stable} and will now prove to be  essential.

Let us first consider the case of rank one torsion free sub-sheaves. As explained in the Appendix, in this case, it is enough to consider all line bundles $L$ such that
\be
L \subset V.
\ee
Using the ample class $H$ in (\ref{s.6}), the slope of $L$ is given by
\be\label{5.4}
\mu_{H}(L)=c_1(L)\cdot H^2=c_1(L)\cdot(H_0^2+2nH_0\cdot H^{'}+n^2h^{'2}f).
\ee
Using the fact that $h^{'2}>0$, and from assumption  (\ref{5.3}), we have
\be
c_1(L|_f)=c_1(L)\cdot f <0.
\ee
It follows that the last term in (\ref{5.4}) is negative. If one can show that 
\be
c_1(L)\cdot H_0^2,\;\;\;\;\;c_1(L)\cdot H_0\cdot H^{'}
\ee
have an upper bound for all line bundles $L$, then there will always exist  a  positive integer $n_0$ such that the slope of every sub-sheaf $L \subset V$ is negative for all $n\geq n_0$.

To prove that this is indeed the case, we will use a result of  \cite{fr} which states that for any given rank four vector bundle $V$ there is an filtration,
\be\label{s.8}
0\subset F^0\subset F^1\subset F^2\subset F^3=V,
\ee
of  vector bundles $F^i,\;i=0,1,2,3$ of rank $i+1$ such that the quotients $F^j/F^{j-1},\;j=1,2,3$ are torsion free rank one sheaves. This implies that these  quotients are of the form
\be
F^j/F^{j-1}\cong L_j\otimes I_{X_j},\;j=1,2,3
\ee 
where the $L_j$ are line bundles and  $I_{X_j}$ are ideal sheaves on $X$  whose sections vanish on $X_j \subset X$, where $X_j$ are at least of codimension two. This, implies that
\be\label{s.7}
c_1(I_{X_j})=0,\;j=1,2,3.
\ee 
What is the image of $L$ in $F^2$? If it is zero, then we obtain a map 
\be\label{5.7}
L \to F^3/F^{2}\cong L_3\otimes I_{X_3}.
\ee
If it is non-zero, then,  using  arguments  similar to the above, $L$ injects into the rank two bundle $F^{2}$. One can now proceed by induction, proving that\be\label{5.5}
L \to F^j/F^{j-1}\cong L_j\otimes I_{X_j}
\ee
for some fixed $j$. Denote this fixed value of $j$ by $J$. Let us consider the map (\ref{5.5}) more closely. It clearly induces a short exact sequence
\be\label{5.6}
\ses{L}{L_J\otimes I_{X_J}}{Q_J},
\ee
where we have defined 
\be
Q_J\equiv (L_J\otimes I_{X_J})/L
\ee
However, since both $L$ and $L_J\otimes I_{X_J}$ are of rank one, $Q_J$ is of rank zero and, hence,  torsion. Under these circumstances,  there always exists an effective divisor $E$ \cite{fr} such that the quotient 
\be
\tilde{Q}_J\equiv (L_J\otimes I_{X_J})/(L\otimes \ox(-E))
\ee
is torsion free. Since the quotient $\tilde{Q}_J$ was torsion to begin with, it must vanish. Hence, we obtain
\be\label{ex039}
L\otimes \ox(-E)\cong L_J\otimes I_{X_J}
\ee
for some effective divisor $E$ and some fixed $J$. Since $E$ and $H^{'}$  are effective and $H_0$ is ample, we obtain the following restrictions on the intersection numbers
\be\label{ex038}
E \cdot H_0^2 > 0,\;\;\;\;E \cdot H_0 \cdot H^{'} > 0.
\ee
Taking the first Chern class of each side of  (\ref{ex039}), and recalling  from (\ref{s.7}) that $ c_1(I_{X_J})$ vanishes, we obtain
\be
c_1(L)-E=c_1(L_J).
\ee
Intersection both sides with   $H_0^{2}$ and $H^{'}\cdot H_0$, and using (\ref{ex038}),
we obtain the inequalities
\be
c_1(L)\cdot H_0^{2} < c_1(L_J)\cdot H_0^{2}
\ee
and 
\be
c_1(L)\cdot H^{'}\cdot H_0 < c_1(L_J)\cdot H^{'}\cdot H_0.
\ee
Note that  $L_J$ is  fixed once we have chosen the filtration (\ref{5.7}).  Hence, we have proven that  there  indeed exists an upper bound for each of 
\be
c_1(L)\cdot H_0^2,\;\;\;\;\;c_1(L)\cdot H_0\cdot H^{'}
\ee
for any $L \subset V$. These bounds are given respectively by
\be\label{ex056}
c_1(L_J)\cdot H_0^2,\;\;\;\;\;c_1(L_J)\cdot H_0\cdot H^{'}.
\ee
Note that the index $J$  depends on the chosen line bundle. For any line bundle $L$, we will obtain one fixed $J$ and  the bounds given in  (\ref{ex056}) apply.
Therefore, making $n$ sufficiently large ensures that the slope of any line bundle $L$ with $L\subset V$ and $c_1(L|_f)<0$ does not destabilize $V$.

A similar argument can be applied to any rank two and rank three torsion free sub-sheaf of $V$ whose restriction to the generic fiber has negative degree. To see this, first note from the Appendix that instead of checking the slope of all rank two torsion free sub-sheaves of $V$, we need only to ensure that all line bundles
\be
L \subset \wedge^2 V
\ee
have negative slope. Here, line bundles $L$ whose restriction to a generic fiber $f$ have negative degree correspond to rank two sub-sheaves of $V$ whose restriction to the generic fiber $f$ have negative degree. However,  the rank six vector bundle $\wedge^2 V$ has a  filtration similar to (\ref{s.8}). This filtration will again lead to  upper bounds on 
\be
c_1(L)\cdot H_0^2,\;\;\;\;\;c_1(L)\cdot H_0\cdot H^{'}
\ee
for  all $L \subset \wedge^2 V$. Therefore, an  argument similar to  the above  can be applied to prove that all such line bundles $L$ have negative slope. The case of rank three sub-sheaves of $V$ whose restrictions to a generic fiber have negative degree works in the same way. In this case, one can easily  prove that all line bundles 
\be
L \subset \wedge^3 V,
\ee
whose restrictions to a generic fiber have negative degree, also have  negative slope. This proves the stability of $V$ with respect to all sub-sheaves $F$ where $F|_f$ has strictly negative degree.

However, it much more difficult  to prove that torsion free  sub-sheaves $F$ of $V$, with $F|_f$ of zero degree,  do not destabilize $V$.  This is  beyond the range of the techniques used in this paper.  Therefore, we postpone this proof to a more technical paper \cite{dopr-ii}.  Here, we will simply state the results. We find that such sub-sheaves will not destabilize a generic element $V$ of $Ext^1(V_2,V_1)$ if one imposes the mild restrictions that 
\be\label{ex060}
b_1-a_1>0,\;\;\;\;b_2-a_2>2.
\ee
These  constraints must, therefore, be added to those given in (\ref{4.11}), namely, that $a_i+b_i \in 2{\mathbb Z}, i=1,2$. That is, the elements of a dense, open subset of $Ext^1(V_2,V_1)$ will be stable holomorphic vector bundles if the conditions (\ref{4.11}) and (\ref{ex060}) are satisfied.

\section{Conclusions}\label{conclusions}

Combining the results of the last several sections, let us state our conclusions.  We construct a pair of rank two holomorphic vector bundles $W_i,\;i=1,2$ on $B$ as locally free extensions of rank one torsion free sheaves
\be
\ses{\oy(a_if)}{W_i}{\oy(b_if)\otimes I_{z_{k_i}}},\;i=1,2
\ee 
where
\be\label{9.1}
a_i+b_i \in 2{\mathbb Z},\;\;\;b_1-a_1>0,\;\;\;b_2-a_2>2
\ee
and
\be\label{9.2}
k_1=2,\;\;\;k_2=8.
\ee
On $B^{'}$, we consider line bundles $L_i,\;i=1,2$ with first Chern class of the form
\be
c_1(L_1)=D^{'}-\frac{1}{2}(a_1+b_1)f^{'},\;\;\;c_1(L_2)=-D^{'}-\frac{1}{2}(a_2+b_2)f^{'}
\ee
with 
\be
D^{'}=e^{'}_2+e^{'}_4+e^{'}_6-e^{'}_9-n^{'}_1-n^{'}_2.
\ee
From these  bundles, we create a pair of rank two holomorphic vector bundles $V_i,\;i=1,2$ on $X$ as
\be\label{hll}
V_i=\pi^{'*}W_i \otimes \pi^{*} L_i.
\ee
Finally, we construct rank four holomorphic vector bundles $V$ on $X$ as extensions of $V_2$ by $V_1$. That is,
\be\label{9.4}
\ses{V_1}{V}{V_2}.
\ee
The set of all such rank four bundles $V$ is denoted by $Ext^1_X(V_2,V_1)$ and satisfies \cite{dopr-ii}
\be
\dim_{\mathbb C}Ext^1_X(V_2,V_1)\geq 1.
\ee
The choice of parameters given in (\ref{9.1}) and (\ref{9.2}) guarantees that there is a dense, open subset of $Ext^1_X(V_2,V_1)$ each of whose elements $V$ is stable and satisfies the standard-like model conditions  ${\bf(C1)}$, ${\bf(C2)}$ and  ${\bf(C3)}$. Furthermore, there exists a vector subspace
\be
Ext^1_X(V_2,V_1)^{inv}\subseteq Ext^1_X(V_2,V_1)
\ee
which carries the trivial representation of the ${\mathbb Z}_2\times {\mathbb Z}_2$ automorphism group. This space satisfies the constraint
\be
\dim_{\mathbb C}Ext^1_X(V_2,V_1)^{inv} \geq \frac{1}{4}\dim_{\mathbb C}Ext^1_X(V_2,V_1)
\ee
and, hence, always contains non-trivial extensions. Each element $V$ in $Ext^1_X(V_2,V_1)^{inv}$ is invariant under ${\mathbb Z}_2\times{\mathbb Z}_2 $. Putting everything together, we conclude that there exists a dense, open subset of $Ext^1_X(V_2,V_1)^{inv}$ each of whose elements is a stable, holomorphic rank four bundle which is invariant under ${\mathbb Z}_2\times{\mathbb Z}_2 $ and satisfies the standard-like model conditions ${\bf(C1)}$, ${\bf(C2)}$ and  ${\bf(C3)}$.

We close this paper by discussing  the moduli space of  ${\mathbb Z}_2\times {\mathbb Z}_2$ invariant, rank four  stable holomorphic vector bundles $V$, which we denote by ${\mathcal M}(V)^{inv}$. These moduli can arise from two distinct sources. First, as we have just discussed, they are associated with the dense, open subspace of vector bundles in $Ext^1_X(V_2,V_1)^{inv}$. In the previous section, we have given a lower bound on this part of the moduli space. However, an exact calculation of the number of such moduli is very difficult and will not be presented here. However, there is a second source of vector bundle moduli. The reader should recall from Section~\ref{Vbundles} that the vector bundles $V_1$ and $V_2$ in the defining exact sequence (\ref{ex014}) were each chosen to be at a specific point in  their moduli space. However, varying $V_1$ and $V_2$ within their moduli space will give continuous moduli to $V$. The dimension of this part of the moduli space of $V$ is much easier to compute. It will simply be the sum of the dimensions of the moduli spaces of $V_1$ and $V_2$. This is best illustrated in a specific example. Consider the specific choice  of integers
\be\label{ex063}
b_1-a_1=2,\;\;\;k_1=2
\ee
and
\be\label{ex064}
b_2-a_2=4,\;\;\;k_2=8.
\ee
Note that each obey the conditions (\ref{9.1}) and (\ref{9.2}). From the discussion in Section~\ref{Wbundles} and expressions (\ref{ex080}) and (\ref{ex081}), it follows that $W_1$ and $W_2$ on $B$ have at least  one and three dimensional  moduli spaces respectively.  This implies that the moduli spaces for $\pi^{'*}W_1$ and $\pi^{'*}W_2$ on $X$ are at least one and three dimensional. Since $B^{'}$ and $X$ are simply connected, there are no continuous moduli corresponding to the line bundles $L_i,\;i=1,2$ and their pull-backs on $X$. Hence,
\be
\dim {\mathcal M}(V_1)^{inv}\geq 1,\;\;\;\dim {\mathcal M}(V_2)^{inv}\geq 3.
\ee
Therefore,  from the above discussion, we conclude that  that for the specific choice of integers (\ref{ex063}) and (\ref{ex064})
\be
\dim {\mathcal M}(V)^{inv}> 4.
\ee
A similar argument can always be applied to compute a lower bound on the dimension of ${\mathcal M}(V)^{inv}$ for any choice of parameters in (\ref{9.1}). 

All of the bundles discussed in this conclusion are invariant, but note equivariant, on $X$. This is a consequence of the fact that only such bundles satisfy the particle physics constraints in the context of this paper. However, it is possible that a modification of our approach could lead to equivariant bundles on $X$ consistent with these constraints. Note that $V$ will be equivariant if $V_1$ and $V_2$ are individually equivariant. It is not too hard to show, using (\ref{hll}), that $V_i$ will be equivariant if $1)$ both $W_i$ and $L_i$ are equivariant or $2)$ neither $W_i$ nor $L_i$ is equivariant. Note that the solutions in this paper correspond to equivariant $W_i$ but non-equivariant $L_i$. Condition  $1)$ might be realized by adding a non-trivial bundle to the hidden sector of the theory. This may allow one to choose $L_i$ to be equivariant. On the other hand, condition $2)$ might be satisfied by choosing $W_i \notin Ext^{1}_B(\oy(b_i f)\otimes I_{z_k}, \oy(a_i f))^{inv}$ and, therefore, not equivariant. These options, as well as others, will be discussed elsewhere.

\section*{Acknowledgments}

Ren\'e Reinbacher and Burt Ovrut are supported in part by DOE under contract No. DE-AC02-76-ER-03071 and the NSF Focused Research Grant DMS 0139799. In addition, Ren\'e Reinbacher wishes to acknowledge partial support from an SAS Dissertation Fellowship and a Daimler-Benz Fellowship. Tony Pantev is supported in part by NSF grants DMS 0099715 and DMS 0139799 and the A.P. Sloan Research Fellowship. Ron Donagi is partially supported by  NSF grant DMS 0104354 and by Focused Research Grant DMS 0139799.
We would also like to thank KITP where parts of this work was accomplished.

\appendix

\

\bigskip

\Appendix{Invariant and Equivariant Vector bundles} \label{appendix2}

In this paper, we construct twisted and  untwisted vector
bundles on non-simply connected Calabi-Yau threefolds $Z$. In order to
do so,  we consider Calabi-Yau threefolds $X$ with a fixed point free
${\mathbb Z}_2\times {\mathbb Z}_2$ group action. Hence we obtain a
degree four cover
\be
q_4: X \to Z.
\ee
In order to obtain a vector bundle $V_{Z}$ on $Z$,  we construct vector
bundles $V$ on $X$ which descend to bundles on $Z$. The descent
properties of $V$ are captured in the way $V$ transforms under the
action of the group ${\mathbb Z}_2\times {\mathbb Z}_2$. There are two
possible types of behavior of $V$ relative to this group  action. These
lead to a twisted or untwisted vector bundle  $V_{Z}$ respectively. 

We begin by recalling the general notions of descent, invariance and
equivariance in our context. Suppose that $Y$ is any manifold and $\Gamma$
is a finite group acting freely. Let $M \equiv  Y/\Gamma$ be the quotient manifold and
$q : Y \to M$  the quotient map. For a vector bundle $V$
on $Y$, we say that:
\begin{itemize}
\item $V$  {\em descends} to $M$ if we can find a vector bundle $V_{M}$ 
on $M$ such that  $V$ and $q^{*}V_{M}$ are isomorphic as bundles on
 $Y$;
\item $V$ {\em is $\Gamma$-invariant} if for every element $g \in \Gamma$ we
 have some isomorphism $\phi_{g} : V \widetilde{\to} g^{*}V$;
\item $V$ {\em is $\Gamma$-equivariant} if there is an action of $\Gamma$ on the
total space of $V$ which covers the given action of $\Gamma$ on $Y$ and is
linear on the fibers of $V \to Y$.
\end{itemize}
By descent theory, one knows that in the case of a free $\Gamma$-action
(which is the only case we will be looking at) the equivariance of $V$
implies that $V$ descends to $M$. In fact, when $V$ is equivariant, the
descended bundle $V_{M}$ is simply the quotient $V_{M} =
V/\Gamma$. Also, from the definitions it is clear that every
$\Gamma$-equivariant $V$ is also $\Gamma$-invariant. In general, the
property of $V$ being $\Gamma$-equivariant is strictly stronger than
the property of $V$ being $\Gamma$-invariant. To compare these two
notions, it is useful to observe that an equivariant structure on $V$
is the same thing as a choice of vector bundle isomorphisms $\phi_{g}
: V \to g^{*}V$ for all $g \in \Gamma$, satisfying the extra
constraint that 
\be
g^{*}(\phi_{h})\circ \phi_{g} = \phi_{hg}
\ee
for any two elements $g, h \in \Gamma$.

\begin{ex} Let $E$ be an elliptic curve with an origin $e \in E$ and
let $\Gamma \cong {\mathbb Z}_{2}$. Specifically take $\Gamma = \{ 1,
t_{e'} \}$ for some non-trivial point $e' \in E$ of order two on
$E$. Then $\Gamma$ acts freely on $E$ and we have a double cover map
$q : E \to E'$ to the quotient $E' \equiv  E/\Gamma$ which is another
elliptic curve.

\begin{itemize}
\item[(i)] 
If $L$ is a line bundle of degree one on $E$, then $L$ can not descend
to $E'$ since any line bundle on $E$ which is a pull-back of a line
bundle on $E'$ must have even degree. In particular, $L$ can not be
equivariant for $\Gamma$. It is not hard to check that $L$ is not even
invariant. Indeed, since $\deg L = 1$ it follows that $L ={\mathcal
O}_{E}(p)$ for some point $p \in E$. Thus $t_{e'}^{*}L ={\mathcal
O}_{E}(t_{-e'}(p))$. But $t_{-e'} = t_{e'}$, since $e'$ is of order
two. Also, by the definition of the group law on $E$, the point
$t_{e'}(p)$ is the unique point on $E$ for which
\be
{\mathcal O}_{E}(t_{e'}(p) - e) \cong {\mathcal O}_{E}(p-e)\otimes {\mathcal
O}_{E}(e'-e) = {\mathcal O}_{E}(p + e' -2e).
\ee
Thus $t_{e'}^{*}L \cong {\mathcal O}_{E}(p + e - e') = L\otimes
{\mathcal O}_{E}(e -e')$ and, hence,  $L$  is never invariant. 
\item[(ii)] The same
argument shows that for any line bundle $L$ on $E$ we have
$t_{e'}^{*}L = L \otimes ({\mathcal O}_{E}(e' - e)^{\otimes \deg
L})$. So, if $L$ is any line bundle on $E$ of even degree, then $L$
will be automatically $\Gamma$-invariant and if $L$ is of odd degree,
then $L$ is not $\Gamma$-invariant. In fact, it is easy to see that
line bundles of even degree on $E$ will be $\Gamma$-equivariant. 
Indeed, if $L$ is such a bundle, then we have some isomorphism $\phi :
L \to t_{e'}^{*}L$. The isomorphism $t_{e'}^{*}(\phi)\circ \phi : L
\to L$ must be a multiplication by some constant $a \in {\mathbb
C}^{*}$. Let $b \in  {\mathbb C}^{*}$ be a square root of
$1/a$. Then the isomorphism $b\cdot \phi : L \to t_{e'}^{*}L$
satisfies $t_{e'}^{*}(b\cdot \phi)\circ (b\cdot \phi) =
\operatorname{id}_{L}$ and so constitutes an equivariant structure on $L$.
Clearly, the same argument will show that any $\Gamma$-invariant vector
bundle $V$ with $H^{0}(E,\operatorname{End}(V)) = {\mathbb C}$ will be
$\Gamma$-equivariant.
\item[(iii)] To construct vector bundles that are invariant but not
equivariant, we need to work with more complicated group
actions. Consider the group $\Lambda$ consisting of translations by
points of order two on $E$. In other words $\Lambda = \{ 1, t_{e'},
t_{e''}, t_{e'''}\}$, where $e'$, $e''$ and $e'''$ are the non-trivial
points of order two on $E$. Now $E/\Lambda = E$ and the quotient map 
$q : E \to E$ is just the multiplication by two in the group law on
$E$. 
Consider now a line bundle $L$ of degree two on $E$. By the
argument in (i), we see that $\lambda^{*}L \cong L$ for all $\lambda
\in \Lambda$, that is, $L$ is $\Lambda$-invariant. However $L$ can
never be $\Lambda$-equivariant since the bundles that descend to
$E/\Lambda$ must have a degree divisible by four. 
\end{itemize}
\end{ex}

\

\bigskip

\noindent
As we saw above, a vector bundle $V$ on $Y$ will descend to $M =
Y/\Gamma$ if and only if $V$ is equivariant. We also saw that
bundles that are invariant but not equivariant will not descend to
bundles on $Y/\Gamma$. Since invariance, albeit different, is
nevertheless very close to equivariance, it will be useful to
understand the type of object on $M$ which will descend from  an
invariant, but not equivariant,  bundle on $Y$. This naturally leads to considering bundles
on gerbes,  or equivalently, twisted  bundles on $M$. 

Suppose that $V$ is a $\Gamma$-invariant vector bundle on $Y$. Then we
have a natural group ${\mathcal G}(V)$ associated with $V$ defined as
follows
\begin{itemize}
\item ${\mathcal G}(V)$ consists of all pairs
$(g,\phi)$, where $g \in \Gamma$ and $\phi : V \to g^{*}V$ is a
vector bundle isomorphism,
\item The product of two elements $(g,\phi),
(h,\psi) \in {\mathcal G}(V)$ is given by the rule 
\be(g,\phi)\cdot
(h,\psi) = (gh,h^{*}(\phi)\circ\psi).
\ee
\end{itemize}
Note that the group ${\mathcal G}(V)$ fits in a short exact sequence
of groups
\begin{equation} \label{eq-theta}
1 \to GL(V) \to {\mathcal G}(V) \to \Gamma \to 1,
\end{equation}
where $GL(V)$ is the group of vector bundle automorphisms of
$V$. Furthermore, $V$ will be $\Gamma$-equivariant if there is a group
homomorphism $\Gamma \to {\mathcal G}(V)$ splitting the sequence
\eqref{eq-theta} and the set of all such homomorphisms is in a
one-to-one correspondence with equivariant structures on $V$ or,
equivalently, with bundles on $M$ that pull-back to $V$. 

Now the group ${\mathcal G}(V)$ acts naturally on the total space of
$V$ in a manner compatible with the action of $\Gamma$ on the base. In
other words, if we use the homomorphism ${\mathcal G}(V) \to \Gamma$
to define  a non-faithful action of ${\mathcal G}(V)$ on $X$, then $V$
becomes a ${\mathcal G}(V)$-equivariant bundle on $X$. In particular,
$V$ descends to a vector bundle on the stack quotient $[X/{\mathcal
G}(V)]$ (see \cite{dp} for details). Since the action of ${\mathcal
G}(V)$ on $X$ was defined so that $GL(V)$ stabilizes every point, it
follows that $[X/{\mathcal
G}(V)]$ is a $GL(V)$-gerbe on $M$.  

If,  in addition,  $V$ is a simple bundle, that is if $GL(V) = {\mathbb
C}^{*}$, then $[X/{\mathcal G}(V)]$ is a ${\mathbb
C}^{*}$-gerbe on $M$. Such gerbes are classified by elements in
$H^{2}(M,C^{\infty}({\mathbb C}^{*}))$, where
$C^{\infty}({\mathbb C}^{*})$ denotes the sheaf of germs of
complex nowhere vanishing differentiable functions on $M$. It is well
understood in physics that elements in $H^{2}(M,C^{\infty}({\mathbb
C}^{*})$ classify gauge equivalence classes of $B$-fields on $M$,
so it is not surprising that the bundles on the gerbe $[X/{\mathcal
G}(V)]$ should be related to instantons on $M$ appearing in a
non-trivial $B$-field background. Since $\Gamma$ was assumed to be
finite, we can be even more precise. When $V$ is simple, the extension 
\eqref{eq-theta} becomes the central extension
\[
1 \to {\mathbb C}^{*} \to {\mathcal G}(V) \to \Gamma \to 1
\]
and so is given by a group cohomology element $e_{V} \in
H^{2}(\Gamma,{\mathbb C}^{*})$. But $\Gamma$ is finite and, hence,
this element must come from some cohomology class $e_{V}^{\op{fin}} \in
H^{2}(\Gamma,\boldsymbol{\mu}_{d})$, where $\boldsymbol{\mu}_{d}
\subset {\mathbb C}^{*}$ is the subgroup of $d$-th roots of unity
and $d$ is just the least common multiple of the orders of all elements
in $\Gamma$. In particular we have a sub-extension
\be
\xymatrix@M+3pt@R-8pt{
1 \ar[r] & {\mathbb C}^{*} \ar[r] & {\mathcal G}(V) \ar[r] &
\Gamma \ar[r] & 1 \\
1 \ar[r] & \boldsymbol{\mu}_{d} \ar[r] \ar@{^{(}->}[u] & {\mathcal
G}^{\op{fin}}(V) \ar[r] \ar@{^{(}->}[u] &
\Gamma \ar[r] \ar@{=}[u] & 1
}
\ee
corresponding to the class $e_{V}^{\op{fin}}$. This means that the
${\mathbb C}^{*}$-gerbe 
$[X/{\mathcal G}(V)]$ is induced from the $\boldsymbol{\mu}_{d}$-gerbe 
$[X/{\mathcal G}^{\op{fin}}(V)]$. Alternatively, this property can be
expressed by saying that the class in $H^{2}(M,C^{\infty}({\mathbb
C}^{*}))$ representing the gerbe $[X/{\mathcal G}(V)]$ comes from
a class in $H^{2}(M,C^{\infty}(\boldsymbol{\mu}_{d}))$. Since
$\boldsymbol{\mu}_{d}$ is finite, the sheaf
$C^{\infty}(\boldsymbol{\mu}_{d})$ of smooth maps from $M$ to
$\boldsymbol{\mu}_{d}$ is a subsheaf in the constant sheaf ${\mathbb
C}^{*}$. In particular, the gerbe $[X/{\mathcal G}(V)]$ comes from
a flat $B$-field. 

Vector bundles on ${\mathbb C}^{*}$-gerbes can be described
explicitly in several different ways \cite{dp}. For the purposes of
our discussion, we need a specific description of such bundles using
the formalism of twisted bundles \cite{c}. In this formalism, the data of
specifying a vector bundle on a gerbe on $M$ is replaced by the
equivalent data of  specifying a twisted vector bundle on $M$ with
twisting given by the equivalence class of the gerbe, which is an
element in $H^{2}(M,C^{\infty}({\mathbb C}^{*}))$. Concretely, if
$\alpha \in H^{2}(M,C^{\infty}({\mathbb C}^{*}))$ and if
$_{\alpha}M$ is the corresponding  ${\mathbb C}^{*}$-gerbe, then
a vector bundle $W$ on  $_{\alpha}M$ is the same thing as an $\alpha$-twisted
vector bundle on $M$. The $\alpha$-twisted vector bundles are most
conveniently described via a cocycle representative of
$\alpha$. Choose an open covering $\{ U_{i} \}$ of $M$ such that in
that covering $\alpha$ is represented by a \v{C}ech cocycle $\{
\alpha_{ijk} \}$ with  $\alpha_{ijk} \in H^{0}(U_{ijk},C^{\infty}({\mathbb
C}^{*}))$. Then an $\alpha$-twisted bundle $W$ on $M$ is a
collection $\{ \{W_{i} \}, \{g_{ij} \} \}$, where
\begin{itemize}
\item $W_{i}$ is a vector bundle on $U_{i}$;
\item $g_{ij} : W_{j|U_{ij}} \widetilde{\to} W_{i|U_{ij}}$ is a vector
bundle isomorphism;
\item $g_{ij}$ satisfy the $\alpha$-twisted cocycle condition
\[
g_{ij}g_{jk}g_{ki} = \alpha_{ijk} \cdot \op{id}.
\]
\end{itemize}

Going back to the question of invariance, we can now say that a simple
$\Gamma$-invariant vector bundle $V$ on $X$ descends to an
$\alpha(V)$-twisted bundle $V_{M}$ on $M$, where $\alpha(V) \in
H^{2}(M,C^{\infty}({\mathbb C}^{*}))$ is the class of the gerbe
$[X/{\mathcal G}(V)]$. However, for describing a heterotic string
compactification on $(M,V_{M})$ we need to specify more than the
geometric background of a twisted vector bundle. We need to understand
the gauge fields on twisted bundles and their gauge symmetries. The
natural framework for this is provided by the notion of a twisted
connection. These twisted connections do not depend only on the smooth
gerbe $\alpha(V) \in H^{2}(M,C^{\infty}({\mathbb C}^{*}))$
twisting the vector bundle, but also require the extra data of a full
physical $B$-field whose underlying gerbe is $\alpha(V)$. Recall that
the standard interpretation of a $B$-field as a locally defined two-form
gauge field is equivalent to the mathematical notion of a connection
on a gerbe \cite{cks,hi,s1,s2}. As usual, we are interested only in flat
$B$-fields, that is, flat connections on gerbes, so that we will preserve
supersymmetry. A  gerbe together with a flat connection
fit together in a cocycle representing a class in $H^{2}(M,{\mathbb
C}^{*})$. To spell this out, we resolve the constant sheaf ${\mathbb
C}^{*}$ by the multiplicative de Rham complex
\be
\xymatrix@1{
{\mathbb C}^{*} \ar[r] & C^{\infty}({\mathbb C}^{*})
\ar[r]^-{d\log} & A^{1}_{M,{\mathbb C}} \ar[r]^-{d} &
A^{2}_{M,{\mathbb C}} \ar[r]^-{d} & \cdots .
}
\ee
Now, we have 
\be
H^{2}(M,{\mathbb C}^{*}) = {\mathbb H}^{2}(M, 
\xymatrix@1{C^{\infty}({\mathbb C}^{*})
\ar[r]^-{d\log} & A^{1}_{M,{\mathbb C}} \ar[r]^-{d} &
A^{2}_{M,{\mathbb C}} \ar[r]^-{d} & \cdots} )
\ee
and, hence,  we can represent any element in $H^{2}(M,{\mathbb C}^{*})$
by a hyper-cohomology 2-cocycle \linebreak 
$\{ (\alpha_{ijk},\Lambda_{ij}, B_{i})
\}$, where
$\alpha_{ijk}$ is a nowhere vanishing smooth complex function on
$U_{ijk}$, $\Lambda_{ij}$ is a smooth one form on $U_{ij}$, and
$B_{i}$ is a smooth two-form on $U_{i}$. Furthermore,  $\{
(\alpha_{ijk}, \Lambda_{ij}, B_{i})
\}$  must satisfy the cocycle conditions
\be
\delta \alpha  = 0 ,\;\;\;\delta \Lambda  = d\log \alpha ,\;\;\;\delta B= d\Lambda,\;\;\;d B  = 0,
\ee
where $\delta$ is the usual \v{C}ech coboundary operator.

Thus, a flat $B$-field $\boldsymbol{B}$ comprises the data $\{
(\alpha_{ijk},\Lambda_{ij}, B_{i}) \}$ up to the natural gauge
transformations given by the coboundaries. Also, the $\{ \alpha_{ijk}
\}$ part of the $B$-field gives a cohomology class $\alpha \in
H^{2}(M,C^{\infty}({\mathbb C}^{*}))$ and so determines the gerbe
underlying our $B$-field. Note that in the particular case of the gerbe
$[X/{\mathcal G}(V)]$  coming from an invariant bundle $V$, we have a
natural choice of a gauge equivalence class of flat connections since
this gerbe is induced from the Deligne-Mumford gerbe $[X/{\mathcal
G}^{\op{fin}}(V)]$.

We are now ready to define a twisted connection on the twisted bundle
$V_{M}$. If $\alpha(V)$ is represented by a \v{C}ech cocycle $\{
\alpha_{ijk} \}$ and  $V_{M}$ is represented by $\{ (V_{M})_{i},
g_{ij} \}$,  then an  $\alpha(V)$-twisted connection  on $V_{M}$
is a collection of connections $\boldsymbol{\nabla} = \{ \nabla_{i} \}$, where
\begin{itemize}
\item $\nabla_{i}$ is a connection on the vector bundle $(V_{M})_{i}$,
\item $\{ \nabla_{i} \}$ satisfy the $\boldsymbol{B}$-twisted gauge
invariance condition
\be
\nabla_{j} = g_{ij}^{*} \nabla_{i} - \Lambda_{ij}\cdot \op{id}.
\ee
\end{itemize}
Observe next that
the twisted cocycle condition on $V_{M}$ becomes an ordinary cocycle
condition on $\op{ad}(V_{M})$. Therefore,  $\op{ad}(V_{M})$ is an untwisted
vector bundle on $M$. Combined with the twisted gauge invariance
condition, this gives
\be
\begin{split}
F_{\nabla_{j}} & = \op{ad}_{g_{ij}}F_{\nabla_{i}} - d
\Lambda_{ij}\cdot \op{id} \\
& = \op{ad}_{g_{ij}}F_{\nabla_{i}} - (B_{i} - B_{j})\cdot \op{id}.
\end{split}
\ee
Hence,  $F_{\nabla_{i}} - B_{i}$ is gauge invariant. In other words,
$F_{\boldsymbol{\nabla}} - B$ is a well-defined two-form
with coefficients in the untwisted vector bundle $\op{ad}(V_{M})$.
Therefore, for  a twisted connection we can write a
$\boldsymbol{B}$-corrected hermitian
Yang-Mills equation 
\be
F_{\boldsymbol{\nabla}} - B = 0.
\ee
Following
through the construction of $\boldsymbol{B}$, we see that every
solution to the ordinary hermitian Yang-Mills equation on $V$ gives
rise to a solution of the $\boldsymbol{B}$-twisted hermitian
Yang-Mills equation on $V_{M}$. Therefore the twisted bundle $V_{M}$
will correspond to a heterotic vacuum if and only if $V$ is stable, or at least
poly-stable, on $Y$.

Applying this general discussion to our original problem of
constructing standard model like vacua on $Z$, we see that a rank four
bundle $V$ satisfying the conditions \eqref{ex016} descends to
standard model vacuum $V_{Z}$ on $Z$, if and only if $V$ is stable  and
${\mathbb Z}_2\times {\mathbb Z}_2$-invariant. 
It is also instructive to observe that since $H^{2}({\mathbb
Z}_2\times {\mathbb Z}_2, {\mathbb C}^{*}) = {\mathbb Z}_{2}$,
the obstruction measuring how far a ${\mathbb
Z}_2\times {\mathbb Z}_2$-invariant bundle on $X$ is from being 
equivariant is two-torsion. In particular, if $V$ is any simple
invariant bundle, the bundle $V^{\otimes 2}$ is automatically
equivariant. This simple observation was  used in
Section~\ref{lbundles}.

\

\bigskip

\Appendix{Slope Stability of Vector Bundles}\label{appendix1}

On a smooth projective variety, there is a relatively simple criterion \cite{d,uyau} for  ensuring that a vector bundle  carries a unique connection whose field strength satisfies the hermitian Yang-Mills equation. The vector bundle must be holomorphic, that is, its transition functions have to be holomorphic and it must be stable. The concept of  holomorphic  vector bundles is straightforward. In this section, we  define stability for such bundles. We also give explicit criteria that a holomorphic vector bundle must satisfy in order for it to be stable.

Let us start with  stable holomorphic vector bundles on a smooth complex projective curve, that is,  a Riemann surface  $C$. 
To begin with, we must define the slope $\mu$ of a vector bundle $W$ on a curve $C$. It is defined by
\be
\mu(W)=\frac{c_1(W)}{\rank W}.
\ee
A vector bundle $W$ on $C$ is said to be stable if for each  vector bundle $F$ 
\be
 F \subset W,
\ee
satisfying
\be
\rank(F) < \rank(W),
\ee
the slope condition 
\be\label{2.0}
\mu(F)< \mu(W)
\ee
is obeyed \cite{fr}. 
Having defined stability for vector bundles on a curve $C$, let us consider some examples. The simplest vector bundle is a holomorphic line bundle $L$. When is it stable? Since the rank of a line bundle is one, there is no  line bundle of lower rank and, hence, for line bundles there is nothing to check, all line bundles are stable. Let us now consider rank two vector bundles $W$ on $C$. In this case, we have to check the slope condition for all line bundles  $F \subset W$. That is, we must show that each $F$ satisfies
\be
\mu(F)< \mu(W).
\ee
Before proceeding to higher dimensional varieties, let us consider a specific curve. Take $C\cong \cp{1}$ and $W$ to be any rank two vector bundle of degree zero. Then, it is not to difficult to shown \cite{fr} that $W$ fits into the short exact sequence 
\be\label{a.4}
\ses{\ocp(a)}{W}{\ocp(-a)}
\ee
for some integer $a\geq 0$. Therefore, $W$ is characterized by an element of $Ext^{1}_{\cp{1}}(\ocp(-a),\ocp(a))$. Happily,  we can calculate $Ext^{1}_{\cp{1}}(\ocp(-a),\ocp(a))$. As explained in (\ref{1.20}) of Section~\ref{Wbundles},
\be
Ext^{1}_{\cp{1}}(\ocp(-a),\ocp(a))\cong H^1(\cp{1},\ocp(a)\otimes \ocp(a) ).
\ee
Using Serre duality on $\cp{1}$, we obtain
\be
H^1(\cp{1}, \ocp(2a))\cong H^0(\cp{1}, \ocp(-2a)\otimes K_{\cp{1}})^{*}.
\ee
But $K_{\cp{1}}\cong \ocp(-2)$. Hence, $\ocp(-2a)\otimes K_{\cp{1}}$ has negative degree and  therefore, no global sections. It follows that $H^1(\cp{1}, \ocp(2a)=0$ and there are no non-trivial extensions of $\ocp(-a)$ by $\ocp(a)$. We conclude that the sequence (\ref{a.4}) splits and
\be
W\cong \ocp(-a)\oplus \ocp(a).
\ee
Hence, $W$ has a natural sub-line bundle $\ocp(a)$.  That is,
\be
\ocp(a) \subset W.
\ee
Since $\mu(\ocp(a))=a,\; a\geq 0$ and $\mu(W)=0$, $W$ is never stable. In general, it can be shown \cite{fr} that there are no stable bundles on $\cp{1}$ of any rank.

Although it will not play a role in this paper, it is useful to discuss a somewhat weaker condition on holomorphic vector bundles called semi-stability. In the above example, assume $a=0$. Then,
\be
W\cong \ocp \oplus \ocp
\ee
and, hence 
\be
\ocp \subset W.
\ee
Since both  $\mu(\ocp)$ and $\mu(W)$ vanish, it follows that 
\be
\mu(\ocp)=\mu(W).
\ee
As we show in Section~\ref{Stable}, every other sub-bundle of $\ocp \oplus \ocp$ has slope smaller zero.  $W$ with these properties is called semi-stable. It can be shown \cite{fr} that  all semi-stable bundles on $\cp{1}$ are of the from $\ocp(a)\oplus\ocp(a) $ for some $a$.

Having defined stability for vector bundles on a curve and having given, at least in principle, a criterion to  check it, we now generalize the notion of stability  to algebraic surfaces and threefolds. To begin with,  we must enlarge the space of holomorphic vector bundles to the space of coherent torsion free  sheaves. Coherent sheaves on projective varieties are sheaves, which have a finite resolution by a set of  vector bundles. This is discussed more fully in Section~\ref{Wbundles}. A torsion sheaf $T$ is a coherent sheaf whose fiber is generically zero. More precisely, there are open sets $U\subset C$, such that the only element of $T(U)$ is the zero map. For example, the skyscraper sheaf ${\mathcal O}_p$ defined in Section~\ref{Wbundles} is a torsion sheaf. A torsion free sheaf $F$ is a coherent sheaf which has no sub-sheaves which are torsion sheaves. Clearly, vector bundles, which are locally free, that is, locally isomorphic to some finite sum of the structure sheaf, are torsion free. Therefore, we have the following inclusions
\be
Vect(C)\subset Coh(C)^{\mbox{tf}}\subset Coh(C),
\ee
were $Vect(C)$ denotes the space of all holomorphic vector bundles on $C$, $Coh(C)^{\mbox{tf}}$  the space of all coherent torsion free sheaves on $C$ and $Coh(C)$ the space of all coherent sheaves on $C$.

To calculate the slope of a vector bundle, or any torsion free sheaf, on a projective variety $Y$ of dimension larger than one, we need an ample class $H \in H^2(Y,{\mathbb Z})$. A two-form $H$ is said to be ample, if the intersection of its Poincare dual with every irreducible, effective curve $C\subset Y$ satisfies
\be
H\cdot C >0.
\ee
It can be shown that a complex manifold admits an ample class if and only if it is a projective algebraic variety. However,  it can be shown, that an elliptic fibration  over projective algebraic surfaces is projective. Hence, the Calabi-Yau space $X$ that we are considering in this paper is a projective algebraic variety, and so an ample class $H$ will exist. The slope of a vector bundle, or any torsion free sheaf, on the Calabi-Yau threefold  $X$ is defined with respect to an ample class $H$ as
\be\label{a.6}
\mu_H(V)=\frac{c_1(V)\cdot H^2}{\rank V}.
\ee
Then, $V$ is stable if for all sub-sheaves $F$ of smaller rank,
\be\label{a.7}
\mu_H(F)<\mu_H(V).
\ee
Note that, since all sub-sheaves of torsion free sheaves are torsion free as well, we can restate this, by saying that  $V$ is stable if  all torsion free sub-sheaves of smaller rank satisfy the slope condition (\ref{a.6}).
Now observe that  a  sub-sheaves  having rank  zero implies  that  their generic fiber is zero, hence they are torsion sheaves. Since there are no torsion sub-sheaves of vector bundles, there are no sub-sheaves of rank zero to check. We are left by checking that all rank one, two and three torsion free sub-sheaves $F$ of $V$ obey
\be
\mu_H(F)<0.
\ee 
where we assume  that $c_1(V)=0$.
Let us begin with rank one sub-sheaves $F$. Since $F$ is torsion free, the natural map,
\be\label{a.0}
i: F \to F^{**}
\ee
where $F^{**}$ is the double dual of $F$, is an injection. This follows from the fact, that the kernel of $i$ consist of the torsion part of $F$, which vanishes by assumption.  Since $F^{**}$ is of rank one, it can be shown \cite{fr} that $F^{**}$ is locally free and, hence, is a line bundle. Dualizing the relation  $F \subset V$ twice and using (\ref{a.0}), we find that 
\be
F \subset F^{**} \subset V^{**}\cong V.
\ee
Hence, we have shown that for any rank one sub-sheaf $F\subset W$, we obtain a locally free sub-sheaf of rank one  $F^{**} \subset V$ of $V$. Furthermore, it can be shown  that $\mu(F)=\mu(F^{**})$. It follows that  instead of checking the slope condition for all rank one torsion free sub-sheaves of $V$, one only need to check the slope condition for the subset of locally free rank one sub-sheaves. Now recall that a locally free sheaf of rank one is a line bundle $L$. We conclude that $V$ will be stable with respect to rank one torsion free sub-sheaves if and only if all line bundles 
\be\label{a.1}
L \subset V
\ee
satisfy
\be
\mu(L)<0.
\ee
What about sub-sheaves of higher rank? Assume we have a rank two torsion free sub-sheaf $F$ of $V$. This specifies an  inclusion
\be
0 \to F \to V
\ee
which, in turn, induces a mapping
\be
\wedge^2 F \to \wedge^2 V.
\ee
By definition, $\wedge^2 F$ includes all sections  $s_1\otimes s_2$ of $F\otimes F$ which are of the form $s_1\otimes s_2=-s_2\otimes s_1$.  If $F$ is a rank two vector bundle, then, clearly, the rank of $\wedge^2 F$ is one. This remains  true for rank two torsion free sheaves. Furthermore, since $F$ is torsion free, $\wedge^2 F$ is torsion free \cite{k}. Therefore, by an  argument as  presented in Section~\ref{Stable} to prove (\ref{inc}), the map
\be
\wedge^2 F \to \wedge^2 V
\ee
is an inclusion and, therefore,  $\wedge^2 F$ is a rank one torsion free sub-sheaf of $\wedge^2 V$. Again, that fact that $c_1(F)=c_1(\wedge^2 F)$ for rank two vector bundles \cite{n} continues to hold for rank two torsion free sheaves and allows us to write
\be
\mu(F)=\frac{c_1(F)}{2}=\frac{1}{2}c_1(\wedge^2 F)=\frac{1}{2}\mu(\wedge^2 F).
\ee
Therefore, we only need to show  that
\be
\mu(F)<0
\ee
for all rank one torsion free sub-sheaves of $F$ of $\wedge^2V$. However, by an  argument similar to that  presented above, we need only  check all rank one locally free sub-sheaves of $\wedge^2 V$, that is,  all line bundles $L$
\be\label{a.2}
L \subset \wedge^2 V.
\ee
The case of rank three torsion free sub-sheaves of $V$ can be approached in a similar manner. That is, instead, of checking that the slope of each  rank three torsion free sub-sheaves $F\subset V, $ such that is negative, one only need to show that all line bundles of the form
\be\label{a.3}
L \subset  \wedge^3 V
\ee
have negative slope.  Checking the slope condition 
\be
\mu(L)<0
\ee
for all line bundles satisfying (\ref{a.1}), (\ref{a.2}) and  (\ref{a.3}) greatly simplifies the proof of stability of a rank four vector bundle $V$.

\end{document}